\documentclass[aps,prd,reprint,showpacs,floatfix,longbibliography,nofootinbib,superscriptaddress]{revtex4-1}

\usepackage{tikz,graphicx,booktabs,multirow,array,microtype,mathtools,float}

\usepackage{centernot}
\graphicspath{{figures/}}
\usepackage{enumerate}   
\extrafloats{200}
\usepackage[colorlinks=true,backref=false, linktocpage=true,
citecolor=blue,
urlcolor=blue,linkcolor=blue,
pdfpagemode=UseOutlines]{hyperref}
\newcolumntype{C}{>{$}c<{$}}
\usepackage{dcolumn}

\usepackage{anyfontsize}
\usepackage[shortlabels]{enumitem}

\usepackage{amsfonts}
\AtBeginDocument{
\heavyrulewidth=.08em
\lightrulewidth=.05em
\cmidrulewidth=.03em
\belowrulesep=.65ex
\belowbottomsep=0pt
\aboverulesep=.4ex
\abovetopsep=0pt
\cmidrulesep=\doublerulesep
\cmidrulekern=.5em
\defaultaddspace=.5em
}
\newcommand{\langl}{\begin{picture}(4.5,7)
\put(1.1,2.5){\rotatebox{60}{\line(1,0){5.5}}}
\put(1.1,2.5){\rotatebox{300}{\line(1,0){5.5}}}
\end{picture}}
\newcommand{\rangl}{\begin{picture}(4.5,7)
\put(.9,2.5){\rotatebox{120}{\line(1,0){5.5}}}
\put(.9,2.5){\rotatebox{240}{\line(1,0){5.5}}}
\end{picture}}

\usepackage{amsmath} % wonderful math package
\usepackage{dsfont}  % double strike font
\usepackage{bm}      % bold math

\def\beq{\begin{equation}}
\def\eeq{\end{equation}}
\def\beqs#1\eeqs{\beq\begin{split} #1 \end{split}\eeq}

\long\def\comment#1{}

\makeatletter 
    
\renewcommand\onecolumngrid{% <<<<<<
\do@columngrid{one}{\@ne}%
\def\set@footnotewidth{\onecolumngrid}% <<<<<<<<<<<<<<<<
\def\footnoterule{\kern-6pt\hrule width 1.5in\kern6pt}%
}

\renewcommand\twocolumngrid{% <<<<<<
        \def\footnoterule{% restore rule
        \dimen@\skip\footins\divide\dimen@\thr@@
        \kern-\dimen@\hrule width.5in\kern\dimen@}
        \do@columngrid{mlt}{\tw@}
}%

\makeatother    
\begin{document}

\title{
Neutral pion polarizabilities  from four-point functions in lattice QCD %: \\connected contributions
}
%\author{authors}
%
\author{Frank~X.~Lee}
\email{fxlee@gwu.edu}
\affiliation{Physics Department, The George Washington University, Washington, DC 20052, USA}
\author{Walter Wilcox}
%\email{walter\_wilcox@baylor.edu}
\affiliation{Department of Physics, Baylor University, Waco, Texas 76798, USA}
\author{Andrei~Alexandru}
%\email{aalexan@gwu.edu}
\affiliation{Physics Department, The George Washington University, Washington, DC 20052, USA}
%\affiliation{Department of Physics, University of Maryland, College Park, MD 20742, USA}
%
\author{Chris~Culver}
%\email{c.culver@liverpool.ac.uk}
\affiliation{Department of Mathematical Sciences, University of Liverpool, Liverpool L69 7ZL, United Kingdom}
\author{Shayan Nadeem}
\affiliation{Department of Physics, Baylor University, Waco, Texas 76798, USA}

\begin{abstract}
We report a proof-of-principle lattice QCD  simulation of the electric and magnetic polarizabilities for a neutral pion 
in the  four-point function method. The results are based on the same quenched Wilson ensembles on a $24^3\times 48$ lattice at $\beta=6.0$  with pion mass from 1100 to 370 MeV previously used for a charged pion.
 For electric polarizability, the results are largely consistent with those from the background field method and ChPT. 
 In contrast, there are significant differences for magnetic polarizability among the four-point function method, the background field method, and ChPT. The situation points to the important role of disconnected diagrams for a neutral pion.
We elucidate a transparent quark decomposition in the four-point function method that can be used to shed light on the issue.

\end{abstract}
%\keywords{Suggested keywords}%Use showkeys class option if keyword display desired
\maketitle

%%%%%%%%%%%%%%%%%%%%%%
%\onecolumngrid
\twocolumngrid
%%%%%%%%%%%%%%%%%%%%%%

%%%%%%%%%%%%%%%%%%%%%%%%%%%%%
\section{Introduction}
\label{sec:intro} 

Electromagnetic polarizabilities are fundamental properties that encode information on the internal structure of hadrons. 
Understanding electromagnetic polarizabilities from first principles has been a long-term goal of lattice QCD. 
The standard approach is the background field method which introduces classical static electromagnetic fields to interact with quarks in hadrons.
The appeal of the method lies in its simplicity: only two-point correlation functions are needed to measure the energy shift with or without the external field, which amounts to a standard calculation of  a hadron's  mass. The linear shift is related to dipole moments, and the quadratic shift to polarizabilities. The method is fairly robust and has been widely applied (see~\cite{endrodi2024} for a recent review and a complete list of references).
It has enjoyed the most success for neutral hadrons.
When it comes to charged hadrons, however, the method faces new challenges.
The reason is rather rudimentary: a charged particle accelerates in an electric field and exhibits Landau levels in a magnetic field.  
Such collective motion of the hadron is unrelated to polarizabilities and must be disentangled from the total energy shift in order to isolate the deformation energy on which the polarizabilities are defined. 
The Euclidean two-point correlation function no longer has a single-exponential behavior at large times.
Special techniques have to be developed to analyze such functions for electric fields~\cite{Detmold:2009dx,niyazi2021charged} and magnetic fields~\cite{Bignell_2018,Bignell_2020,Bignell:2020xkf,He:2021eha}.

Partly spurred by the challenges for charged particles, an alternative approach based on four-point functions in lattice QCD has received renewed interest in recent years. 
It offers a transparent physical picture  that treats neutral and charged particles on equal footing; the latter simply having additional elastic contributions in the form of charge radii.
The potential of using four-point functions to access polarizabilities has been investigated in the early days of lattice QCD~\cite{BURKARDT1995441,Andersen:1996qb,Wilcox:1996vx}.
An intermediate method based on a perturbative expansion in the background field at the action level was employed later~\cite{Engelhardt:2007ub}, leading to the same diagrammatic structure as the standard four-functions discussed here.
A reexamination of the formalism in Ref.~\cite{Wilcox:1996vx} was carried out  in Ref.~\cite{Wilcox:2021rtt} in which new formulas were derived in momentum space for electric and magnetic polarizabilities of both a charged pion and the proton.  
Proof-of-principle simulations applying the formulas for charged pion electric polarizability~\cite{Lee:2023rmz} and magnetic polarizability~\cite{Lee:2023lnx} have demonstrated the promise of such methods. 
At the same time, position-space-based four-point function simulations have also emerged for pions~\cite{Feng:2022rkr} and  proton and neutron~\cite{wang2024}. 

In this work, we focus on applying the four-point function method to a neutral pion.
Outside lattice QCD, chiral perturbation theory (ChPT) as constrained by phenomenology provides most solid information on pion polarizabilities~\cite{BURGI1996392,Gasser_2006}. 
At leading order,  ChPT predicts $\alpha_E+\beta_M=0$ for both charged and neutral pions. 
Specifically, $\alpha_E=-\beta_M =3.0$ for a charged pion and $\alpha_E=-\beta_M =-0.5$ for a neutral pion in  standard units of  $10^{-4}\;\text{fm}^3$.
At two-loop order it gives  for a neutral pion,
\beqs
{\bm \pi^0:\qquad} \alpha_E + \beta_M &=1.1 \pm 0.3 \\
\alpha_E - \beta_M &=-1.9  \pm 0.2 \\
\alpha_E &=-0.40 \pm 0.18 \\
\beta_M &=1.5 \pm 0.27.
\label{eq:pizero}
\eeqs
For a charged pion, it gives 
\beqs
{\bm \pi^+: \qquad} \alpha_E + \beta_M &=0.16  \\
\alpha_E - \beta_M &=5.7  \pm 0.1 \\
\alpha_E &=2.93 \pm 0.05 \\
\beta_M &=-2.77 \pm 0.11.
\label{eq:piplus}
\eeqs
We see significant differences (opposite signs) in polarizabilities between a neutral and a charged pion.
This offers a good testing case for the four-point function method on the lattice.
A comprehensive review on pion polarizabilities from non-lattice approaches and experiment 
can be found in Ref.~\cite{Moinester:2019sew}.

In Sec.~\ref{sec:method} we derive the formulas for $\alpha_E$ and $\beta_M$ for a neutral pion and define the associated  four-point functions. 
In Sec.~\ref{sec:results} we  show simulation results, including momentum dependence and pion mass dependence.
In Sec.~\ref{sec:con} we give conclusion and outlook.
Details on the complete four-point correlation functions are given in the Appendix.

%%%%%%%%%%%%%%%%%%%%%%%%%%%%%
\section{Methodology}
\label{sec:method} 

\begin{figure}[h]
\includegraphics[scale=0.45]{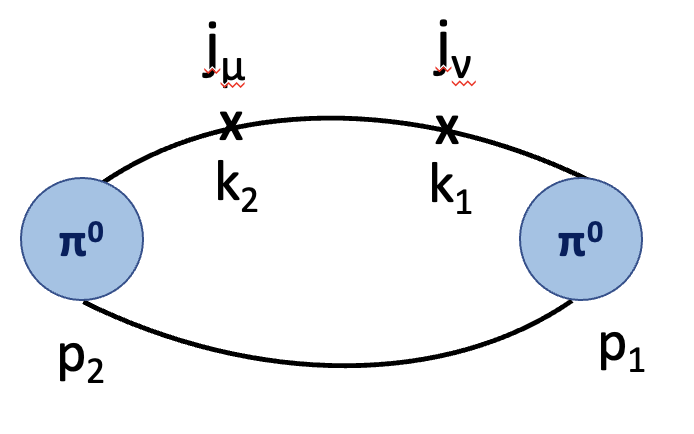}
\caption{Pictorial representation of a four-point function for a neutral pion with quark and anti-quark lines.
The four-momentum conservation is $p_2 +k_2 = k_1+p_1$.}
\label{fig:diagram-4pt1}
\end{figure}

First we outline a derivation of the polarizability formulas for a neutral pion, contrasting them with the ones derived in Ref.~\cite{Wilcox:2021rtt} for a charged pion.
The process is represented in Fig.~\ref{fig:diagram-4pt1}.
Two modifications are needed. One  is in the Compton tensor $T_{\mu\nu}$ which is now absent of the $g_{\mu\nu}$ and $A$ terms,
\beqs
&\sqrt{2E_12E_2} \,T_{\mu\nu} = \\
& -{T_\mu(p_1+k_1,p_1) T_\nu(p_2,p_2+k_2) \over (p_1+k_1)^2-m^2 } \\
& -{T_\mu(p_2,p_2-k_1) T_\nu(p_1-k_2,p_1) \over (p_1-k_2)^2-m^2 } \\ %+2g_{\mu\nu} \\
%&+ A(k_1^2g_{\mu\nu} - k_{1\mu}k_{1\nu} + k_2^2 g_{\mu\nu} - k_{2\mu}k_{2\nu}) \\
&+ B(k_1\cdot k_2 g_{\mu\nu} -  k_{2\mu}k_{1\nu}) \\
&+ C(k_1\cdot k_2 Q_\mu Q_\nu + Q\cdot k_1 Q\cdot k_2  g_{\mu\nu} \\
&\quad\quad - Q\cdot k_2 Q_\mu k_{1\nu} - Q\cdot k_1 Q_\nu k_{2\mu}),
\label{eq:Tmn}
\eeqs 
where 
\beq
B={2m_\pi\beta_M\over \alpha},\;
 C= -{\alpha_E +\beta_M \over 2m_\pi\alpha}\label{eq:BC}
\eeq
Here $\alpha=1/137$ is the fine structure constant.
The tensor  still obeys the current conservation condition,
\beq
 k^\mu_1T_{\mu\nu}= k^\nu_2T_{\mu\nu}=0,
 \eeq
 and has the expected low-energy expansion that defines the polarizabilities,
\fontsize{9}{9}
\beqs
\alpha \epsilon^{\mu}_1 T_{\mu\nu} \epsilon^{\nu*}_2 = 
\alpha_E\, \omega_1\omega_2\hat{\epsilon}_1\cdot \hat{\epsilon}_2^* 
+ \beta_M\, (\hat{\epsilon}_1\times \vec{k}_1)\cdot (\hat{\epsilon}^*_2\times \vec{k}_2), 
\label{eq:low}
\eeqs
\normalsize
where $\epsilon_1$ and $\epsilon_2$ are the initial and final photon polarization 4-vectors.

The other modification is in the electromagnetic  vertex function in the Born part of the tensor (first two terms in Eq.\eqref{eq:Tmn}).
%
%\beq
%T_\mu(p',p)=(p'_\mu+p_\mu) F_\pi(q^2) +q_\mu {p'^2-p^2\over q^2}(1-F_\pi(q^2)),
%\eeq
%%
%where  $q=p'-p$  is the momentum transfer.
%It satisfies $q_\mu T_\mu(p',p)=p'^2-p^2$ for off-shell pions, which is needed for the Ward-Takahashi identity.
Since the $\pi^0$  form factor $F_\pi(q^2) $  vanishes in the isospin limit, there is no Born contribution in the tensor.

\begin{figure}[h]
\includegraphics[scale=0.4]{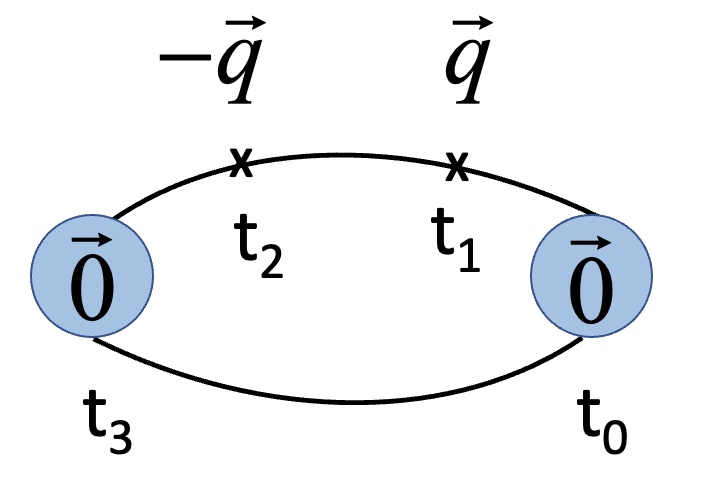}
\caption{Zero-momentum Breit frame in Euclidean space. 
The four-momentum conservation is recast as $(m_\pi,\vec{0} )= (0,-\vec{q})+(0,\vec{q})+(m_\pi,\vec{0})$  on the lattice. The four time points are placed at $t_0$ (source), $t_1$ (current 1), $t_2$ (current 2), $t_3$ (sink).}
\label{fig:diagram-4pt2}
\end{figure}

Then a matching procedure in momentum space of the continuum tensor and its lattice version in the zero-momentum Breit frame (see Fig.~\ref{fig:diagram-4pt2}) leads to the following formula,
\beq
\alpha_E
= \lim_{\bm q\to 0}{2\alpha \over \bm q^{\,2}} \int_{0}^\infty d t \;Q_{44}(\bm q,t), 
 \label{eq:alpha} 
\eeq
for electric polarizability, and
\beq
\beta_M
=\lim_{\bm q\to 0}{2\alpha \over \bm q^{\,2}} \int_{0}^\infty d t \bigg[Q_{11}(\bm q,t) -Q_{11}(\bm 0,t) \bigg],
 \label{eq:beta} 
\eeq
for magnetic polarizability.
The formulas are in discrete Euclidean spacetime but we keep the time axis continuous for notational convenience.
%Zero-momentum Breit frame is employed in the formula to mimic low-energy Compton scattering, where the initial and final   pions are at rest and the photons have purely spacelike momentum.
%

Comparing with the formulas for a charged pion~\cite{Wilcox:2021rtt}, there are a number of differences.
For $\alpha_E$, there is no elastic contribution for $\pi_0$. Both the charge radius term  $\alpha \langl r_E^2\rangl /(3m_\pi)$ and $Q_{44}^{elas}$ vanish, leading to a much simpler formula.
The sign of $\alpha_E$ for $\pi^0$ is directly given by the sign of the time integral over $Q_{44}$.
For $\beta_M$, the charge radius term is also absent,  making its sign solely dependent on the sign of the  subtracted time integral  over $Q_{11}$. 
Since there is no elastic contributions for $\pi^0$, we remove the redundant  `inel' label from $Q_{11}^{inel}$ that is used in the formula for $\pi^+$.
%The zero-momentum sum rule becomes $T_{11}(\bm 0)=0$  for $\pi^0$  instead of $T_{11}^{inel}(\bm 0)={-1\over m_\pi}$  for $\pi^+$.
%This means any nonzero signal can be attributed to disconnected insertions since we only consider connected insertions in this work.
Finally, the four-point correlation function  at the quark level will be different due to the different interpolating field for a neutral pion.
The four-point function is defined on the lattice by ($\mu=1$ and $4$ in this work),
\fontsize{9.5}{9.5}
\begin{align}
&Q_{\mu\mu}(\bm q,t_3,t_2,t_1,t_0)  \equiv  \label{eq:Q11} \\&
 \frac{ \displaystyle\sum_{\bm x_3,\bm x_2,\bm x_1,\bm x_0} e^{-i\bm q\cdot \bm x_2} e^{i\bm q\cdot \bm x_1} 
\langl \Omega | \psi (x_3) :j^L_\mu(x_2) j^L_\mu(x_1):  \psi^\dagger (x_0) |\Omega \rangl }
{\displaystyle\sum_{\bm x_3,\bm x_0} \langl \Omega  | \psi (x_3) \psi^\dagger (x_0) |\Omega \rangl }. \nonumber
\end{align}
  \normalsize
The interpolating field for a neutral pion is given by,
\beq
\psi_{\pi^0}(x)={1\over \sqrt{2}} \left[\bar{u}(x) \gamma_5 u (x)-\bar{d}(x) \gamma_5 d (x)\right ].
\label{eq:op}
\eeq
The resulting correlation function has self-contracting quark loops at the source and sink whereas that of a charged pion  %$\bar{d}(x) \gamma_5 u (x)$ 
does not. 

%In this expression, $\Omega$ denotes the vacuum, and normal ordering is used to formally include the required subtraction of vacuum expectation values (VEV) on the lattice, which comes in the form of $ \langl \Omega | T j^L_\mu(x_2) j^L_\mu(x_1) | \Omega\rangl$. 
%The sums over $\bm x_0$ and $\bm x_3$ enforce zero-momentum pions at the source ($t_0$) and sink ($t_3$). 
%The sum over $\bm x_1$ injects momentum $\bm q$ by the current at $t_1$, 
%whereas sum over $\bm x_2$ takes out $\bm q$ by the current at $t_2$ to satisfy energy-momentum conservation in the process.
%The two possibilities of time ordering are implied in the normal ordering.

 For the lattice version of electromagnetic current density $j^L_\mu$,
we consider two options. One is a local current (or Point Current) built from up and down quark fields,
\beqs
j^{(PC)}_\mu &\equiv f\,\,Z_V \kappa\left(q_u \bar{u}\gamma_\mu u + q_d \bar{d}\gamma_\mu d \right) \\
 f&=\{1,i\}  \text{ for }  \mu=\{4,1\} .
\label{eq:j1PC}
\eeqs
The extra  factor  of $i$ in the magnetic case is needed  to ensure that the spatial component $j^{(PC)}_1$ is hermitian.
The reason is that $(\bar{u}\gamma_1 u)^\dagger=- \bar{u}\gamma_1 u$  whereas  $(\bar{u}\gamma_4 u)^\dagger=\bar{u}\gamma_4 u$ (recall $\bar{u}\equiv u^\dagger \gamma_4$).
The factor $\kappa$ is to account for the quark-field rescaling $\psi\to \sqrt{2\kappa} \psi$ in Wilson fermions. The factor of 2 is canceled by the 1/2 factor in the definition of the vector current ${1\over 2}\bar{\psi}\gamma_\mu \psi$.
The charge factors are $q_u=2/3$ and $q_d=-1/3$ where the resulting $e^2 =4\pi  \alpha$ (in the unit system of  $\hbar=c=\epsilon_0=1$) in the four-point function has been absorbed in the definition of  $\alpha_E$ in  Eq.\eqref{eq:alpha} and $\beta_M$ in  Eq.\eqref{eq:beta}. 
The advantage of the local operator is that it leads to relatively simple correlation functions. 
The drawback is the issue of renormalization constant $Z_V$ for vector current on the lattice. 
The other option is the conserved vector current (or Point-Split current) for Wilson fermions,
\beqs
 &j^{(PS)}_\mu (x) \equiv \\&
\quad \,f\, q_u {\kappa_u} \big[ 
-\bar{u}(x) (1-\gamma_\mu) U_\mu(x) u(x+\hat{\mu}) \\& \qquad\qquad
+
\bar{u}(x+\hat{\mu}) (1+\gamma_\mu ) U_\mu^\dagger(x) u(x) 
\big] 
\\ &
+  f\,\,q_d {\kappa_d} \big[ 
-\bar{d}(x) (1-\gamma_\mu) U_\mu(x) d(x+\hat{\mu}) \\&  \qquad \qquad
+ 
\bar{d}(x+\hat{\mu}) (1+\gamma_\mu ) U_\mu^\dagger(x) d(x) 
\big].
\label{eq:j1PS}
\eeqs
%
%The same factor $f$ applies here.
Although the conserved current explicitly involves gauge fields and leads to more complicated correlation functions, it has the advantage of circumventing the renormalization issue ($Z_V\equiv 1$). All numerical results in this work  are based on conserved current. 

\begin{figure}[thb!]
\includegraphics[scale=0.45]{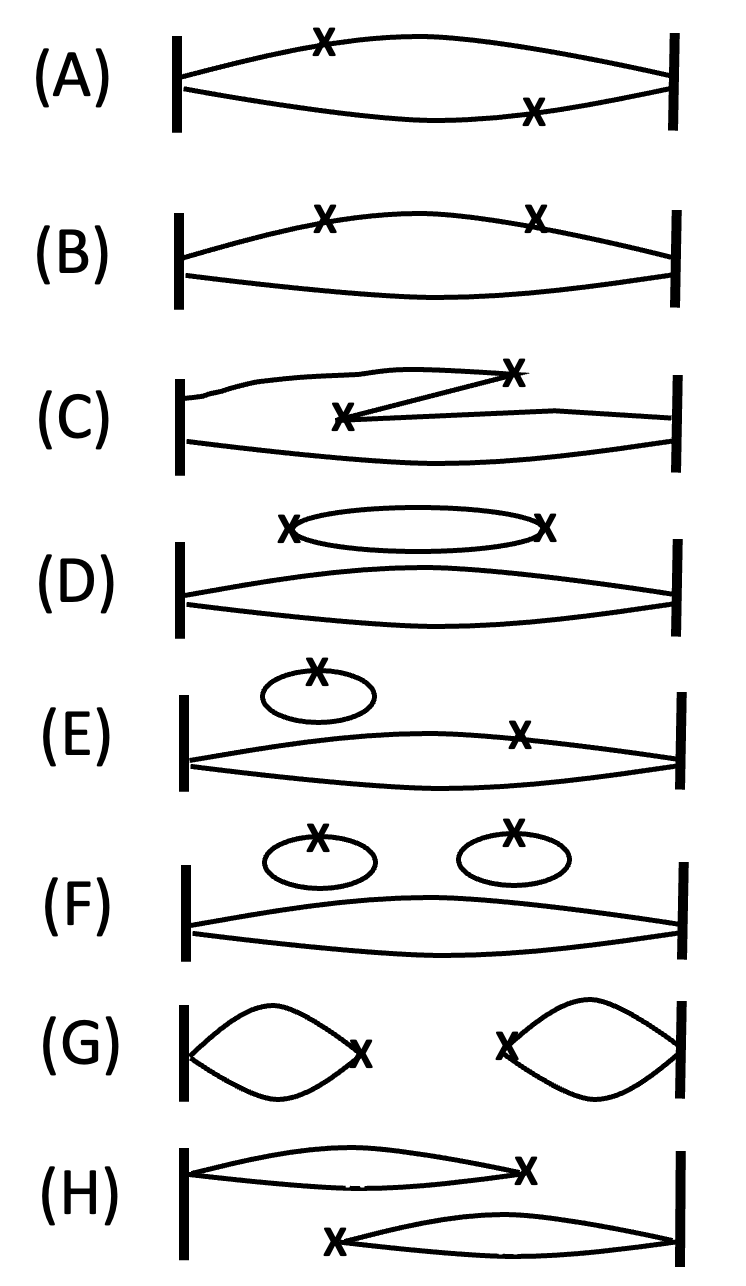}
\caption{Quark-line  diagrams of a four-point function contributing to polarizabilities of a neutral pion.
% (a) connected insertion: different flavor, (b) connected insertion: same flavor, (c) connected insertion: same flavor Z-graph, (d) disconnected insertion: single loop, double current, (e) disconnected insertion: single loop, (f) disconnected insertion: double loop.  
Each diagram represents a distinct topology, with flavor and additional time orderings assumed as well as gluon lines that connect the quark lines. Current insertions are represented by crosses.  Zero-momentum pion interpolating fields are represented by vertical bars (wall sources).}
\label{fig:4pt}
\end{figure}

Wick contractions of quark-antiquark pairs in $Q_{\mu\mu}$ in  Eq.\eqref{eq:Q11} lead to topologically 
distinct quark-line diagrams shown in Fig.~\ref{fig:4pt}. Compared to diagrams for a charged pion~\cite{Lee:2023rmz,Lee:2023lnx}, two new disconnected diagrams (G and H) emerge for a neutral pion.
They are responsible for the leading $1/m_\pi$ behavior in neutral pion polarizabilities with a model-independent coefficient in ChPT~\cite{He2020}. 
The complete expressions are given in the Appendix where we point out relations in various parts of the correlation functions. Note that diagram E represents four possibilities in flavor and time orderings in Eq.\eqref{eq:all}.
The total contribution is simply the algebraic sum of the normalized individual terms,
\beqs
Q_{\mu\mu}(\bm q,t_2,t_1)&=\sum_{k=A,B,C,D,E,F,G,H} Q^{(k)}_{\mu\mu}.
\label{eq:Q11abc}
\eeqs
It holds for either local current or conserved current. The charge factors and flavor-equivalent contributions have been included in each diagram. One can examine the diagrams one by one, building a transparent physical picture for the polarizabilities.
For numerical results, we focus on the connected contributions (diagrams A,B,C) in the isospin limit ($\kappa_u=\kappa_d$) in this study. 
The disconnected contributions (diagrams D,E,F,G,H) are more challenging and are left for future work.

%

%%%%%%%%%%%%%%%%%%%%%%%%%%%%%%%%%%%%%%%%%%%%%%%%%%%%%%%%%%%%%%%%%%%%
\section{Simulation details and results}
\label{sec:results}
%%%%%%%%%%%%%%%%%%%%%%%%%%%%%%%%%%%%%%%%%%%%%%%%%%%%%%%%%%%%%%%%%%%%
As a proof-of-principle test, we use quenched Wilson action with $\beta=6.0$ and $\kappa=0.1520,\;0.1543,\; 0.1555,\; 0.1565$ on the lattice $24^3\times 48$.
The corresponding pion masses are $m_\pi=1100, 800, 600, 370$ MeV.
We analyzed 1000 configurations for each of the kappas.
%The scale of this action has been determined in Ref.~\cite{CABASINO1991195}, with inverse lattice spacing $1/a=2.312$ GeV and kappa critical $\kappa_c=0.15708$.
%It also gives the pion mass as a function of kappa,
%\beq
%(m_\pi a)^2 = 2.09 \times \frac{1}{2}\bigg(\frac{1}{\kappa} - \frac{1}{\kappa_c}\bigg).
%\label{eq:scale}
%\eeq
Dirichlet (or open) boundary condition is imposed in the time direction where the fermion field is set to zero on the time boundary via gauge links in the hopping terms of the fermion action. Periodic boundary conditions are used in spatial dimensions.
The pion source is placed at $t_0=7$ and sink at $t_3=42$ (time is labeled from 1 to 48). One current is inserted at a fixed time $t_1$, while the other current $t_2$ is free to vary.
%We use integers $\{n_x,n_y,n_z\}$ to label the discrete momentum on the lattice,
%\beqs
% \bm q &=\big\{{2\pi n_x\over L_x}, {2\pi n_y\over L_y}, {2\pi n_z\over L_z} \big\}, \\& 
% \quad n_x,n_y,n_z =0, \pm 1, \pm 2, \cdots,
% \label{eq:q}
%\eeqs
We consider five different combinations of momentum $\bm q=\{0,0,0\},\,\{0,0,1\},\,\{0,1,1\},\{1,1,1\},\,\{0,0,2\}$. 
In lattice units they correspond to  the values ${\bm q}^2a^2=0,\, 0.068,\, 0.137,\, 0.206,\, 0.274$, 
or in physical units to ${\bm q}^2=0,\, 0.366,\, 0.733,\, 1.100,\, 1.465$ (GeV$^2$).
%Note that  $\{1,1,1\}$ is not excluded since there is no elastic contributions in $\beta_M$ for $\pi^0$. 

%%%%%%%%%%%%%%%%%%%%%%%%
\subsection{Raw correlation functions}
\begin{figure*}[htb!]
\includegraphics[scale=0.47]{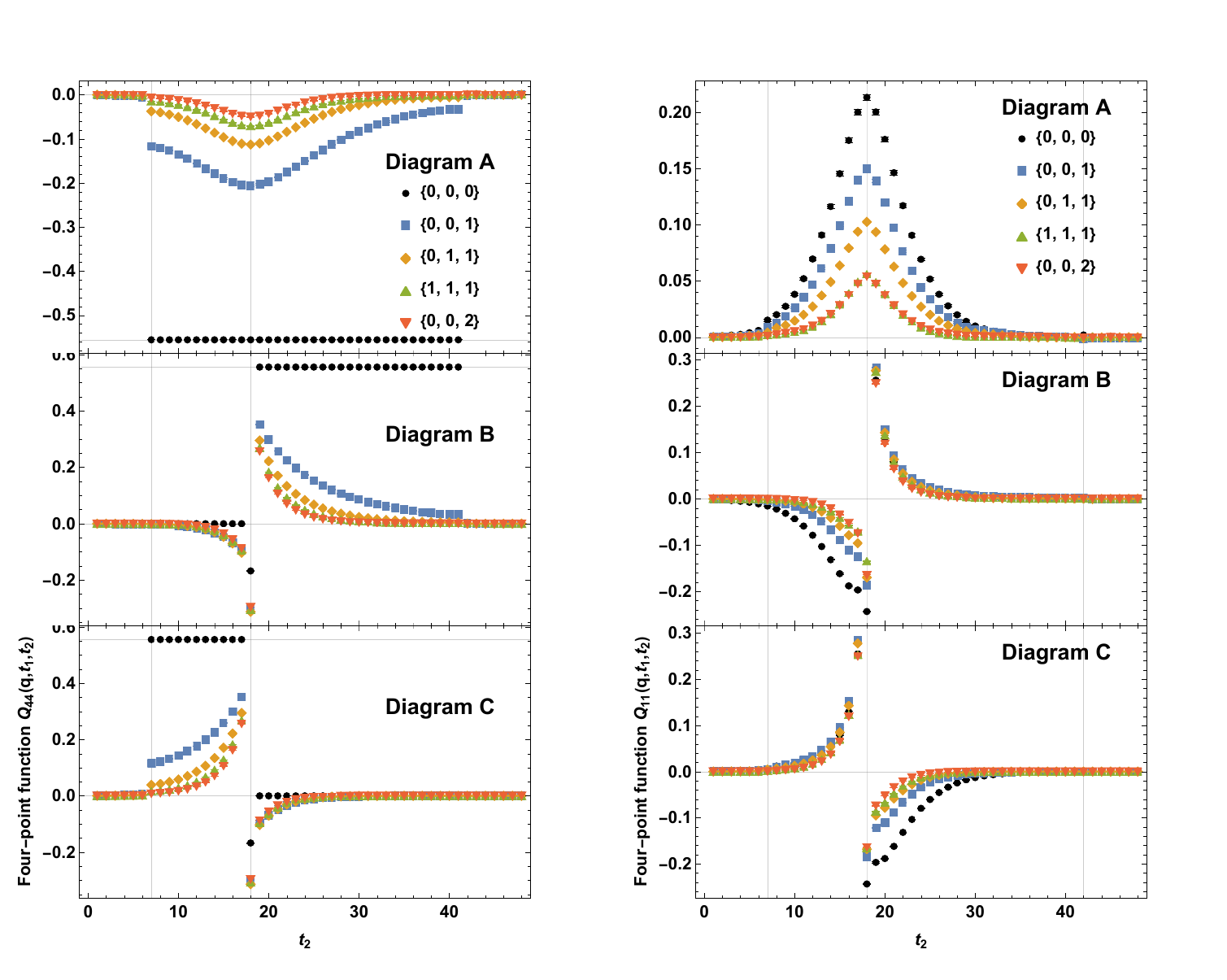}
\caption{
Individual four-point functions $Q_{44}$ (left panel) and $Q_{11}$ (right panel) from the connected diagrams at different momentum and $m_\pi=600$ MeV. 
Vertical gridlines indicate the pion walls ($t_0=7$ and $t_3=42$) and the  fixed current insertion ($t_1=18$).
Horizontal gridlines in $Q_{44}$ indicate charge conservation factor of $-10/9$ for diagram A and $10/9$ for diagrams B and C.
The results between $t_2=19$ and $t_2=41$ will be the basis for neutral pion polarizabilities.
}
\label{fig:Q11PS}
\end{figure*}
\begin{figure*}[htb!]
\includegraphics[scale=0.47]{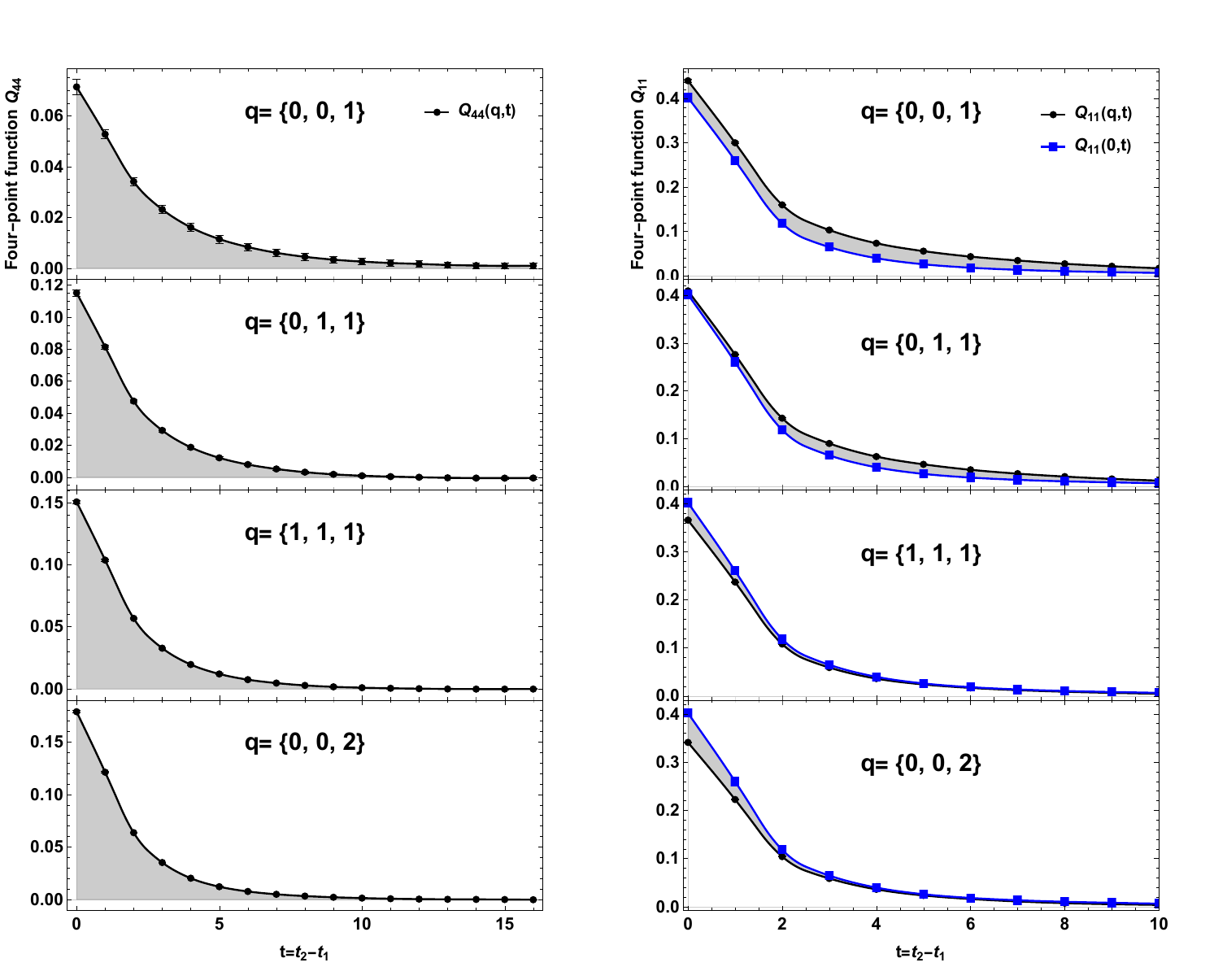}
\caption{
Connected contributions to $Q_{44}(\bm q,t)$ (left panel) and  $Q_{11}(\bm q,t)$ (right panel)  at different values of $\bm q$ and $m_\pi=600$ MeV.
The shaded areas are the dimensionless signal contributing to $\pi^0$ polarizabilities.
}
\label{fig:QQ}
\end{figure*}

In Fig.~\ref{fig:Q11PS} we show the raw normalized four-point functions from the leading-order connected diagrams,
 both electric and magnetic,  at the five different values of momentum $\bm q$ and at $m_\pi=600$ MeV. 
 %They are based on conserved current.
All points are included and displayed on a linear scale for comparison purposes.

For the electric $Q_{44}$, charge conservation at zero momentum (or charge-charge overlap) leads to a simple relation,
\fontsize{9.5}{9.5}
\begin{align}
 \frac{ \displaystyle\sum_{\bm x_3,\bm x_2,\bm x_1,\bm x_0} 
\langl \Omega | \psi (x_3) j^{(PS)}_4(x_2) j^{(PS)}_4(x_1)  \psi^\dagger (x_0) |\Omega \rangl }
{\displaystyle\sum_{\bm x_3,\bm x_0} \langl \Omega  | \psi (x_3) \psi^\dagger (x_0) |\Omega \rangl }=q_1q_2,
\label{eq:cc}
\end{align}
where $q_1$ and $q_2$ are the charge factors of the two quarks that the two currents couple to. It provides a powerful check on the implementation of conserved current on the lattice. Indeed, we see the factor $2(q_uq_{\bar{u}}+q_dq_{\bar{d}})=-10/9$ exactly satisfied (black dots)  in diagram A where the two currents couple to different quark lines. The factor of 2 comes from two equivalent flavor contributions in the isospin limit. 
The charge factor is $2(q_uq_u+q_dq_d)=2(q_{\bar{u}}q_{\bar{u}}+q_{\bar{d}}q_{\bar{d}})=+10/9$  for diagram B and C where the two currents couple to the same quark line.

The special point of $t_1=t_2$ is regular in diagram a when they couple to different quarks, but gives irregular results in diagram B and C when they couple to the same quark.  They are present in both $Q_{44}$ and $Q_{11}$ and at all values of $\bm q$.  The irregularity is due to unphysical contact interactions on the lattice which vanish in the continuum limit. We handle this point with special care in our analysis below. 

The results about $t_1=18$ in diagram B and C are mirror images of each other,  simply due to the fact that they are from the two different  time orderings of the same diagram. In principle, this property could be exploited to reduce the cost of simulations by placing $t_1$ in the center of the lattice. In this study, however, we computed all three diagrams separately, and add them between $t_1=19$ and $t_3=41$ as the signal.

%%%%%%%%%%%%%%%%%%%%%%%%
\subsection{Polarizabilities}
\begin{figure}[bht!]
\includegraphics[scale=0.4]{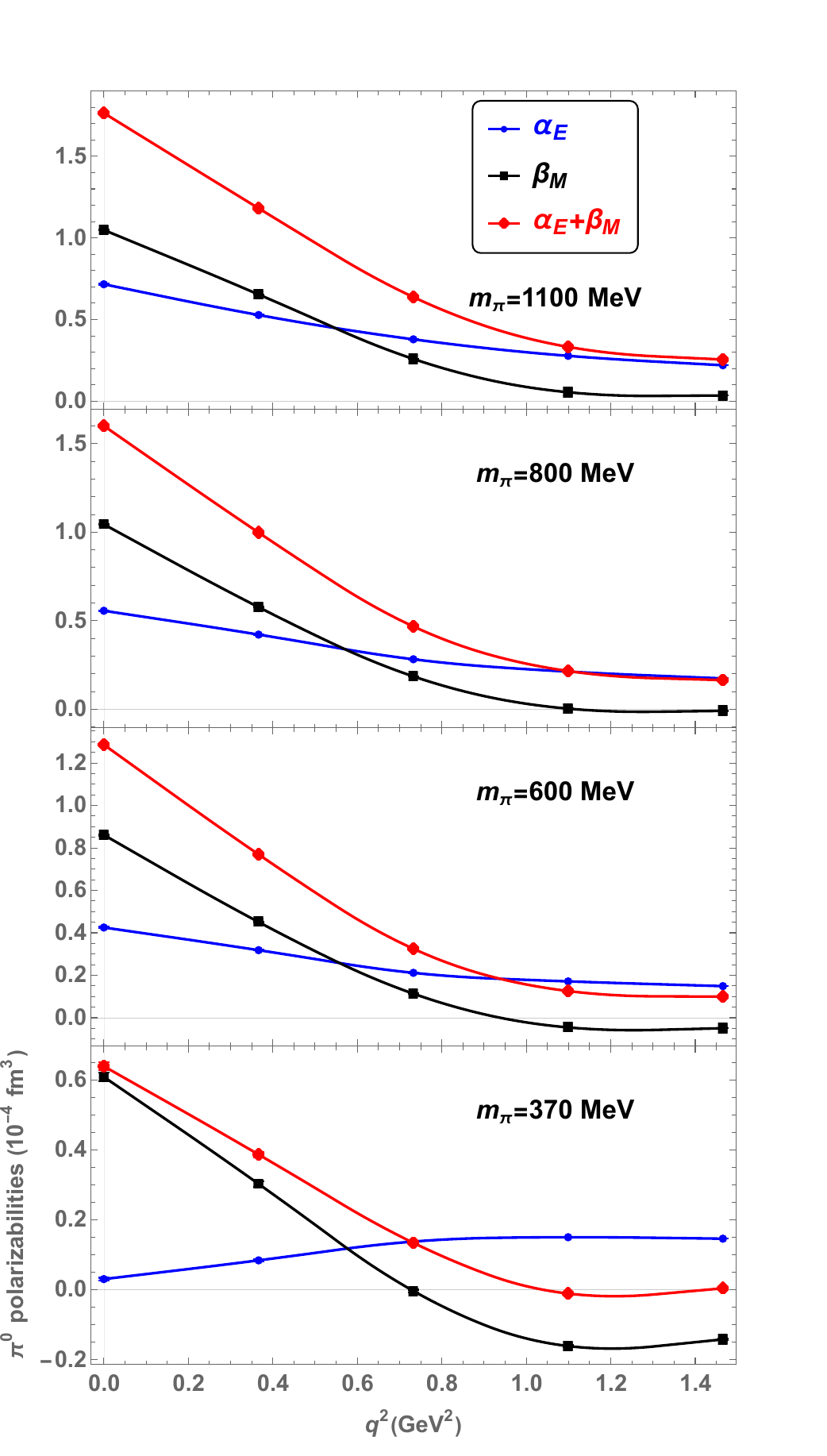}
\caption{
Momentum dependence of neutral pion polarizabilities at different pion masses.
The static values are obtained by linear extrapolation to $\bm q^2=0$  using the two lowest $\bm q^2$ points with lines connecting all points smoothly.
}
\label{fig:Qzero}
\end{figure}

In Fig.~\ref{fig:QQ} we show in lattice units  the connected contribution $Q_{44}$ and $Q_{11}$ at different $\bm q$ values as a function of current separation $t=t_2-t_1$.  Only results for $m_\pi=600$ MeV are shown as an example; the graphs at the other pion masses look similar.  
The time integrals in the formulas Eq.\eqref{eq:alpha} and Eq.\eqref{eq:beta} for $\pi^0$ polarizabilities are given by the shaded areas.
For  $Q_{44}$, the area is under a single curve and is positive at all ${\bm q}^2$ values. For $Q_{11}$, the area is from the difference between $Q_{11}(\bm q)$ and $Q_{11}(0)$ curves, and there is a switch over of the two curves, indicating a sign change from positive to negative. 
Note the difference plotting scales in left and right panels: the signal is on the same order of magnitude.
One detail to notice is that the curves include the $t=0$ point which is the unphysical contact term mentioned earlier. We would normally avoid this point and only start the integral from $t=1$. However, the chunk of area between $t=0$ and $t=1$ is the largest piece in the integral. To account for  this contribution, we linearly extrapolated both  $Q_{44}$  and  $Q_{11}$ back to $t=0$ using the two points at $t=1$ and $t=2$. As the continuum limit is approached, the $t=0$ point will become regular and the chunk will shrink to zero.

To obtain polarizabilities, we multiply the shaded area by the factor $2\alpha/{\bm q}^2$ and convert to physical units. Since $\alpha_E$ and $\beta_M$ are static properties, we take the limit at $\bm q^2=0$ by a simple linear extrapolation using the two lowest points. The results are displayed in Fig.~\ref{fig:Qzero} for all pion masses. The $\alpha_E$ (blue)  remains positive at all $\bm q^2$ values but the extrapolated value decreases as smaller pion masses. The  $\beta_M$ (black) has a sign change and the extrapolated value decreases as pion mass decreases.  The $\bm q^2$ dependence of $\alpha_E$ and $\beta_M$ crosses over at around 0.58 GeV$^2$.
The sum of electric and magnetic polarizabilities $\alpha_E+\beta_M$ has smooth $\bm q^2$ dependence with a decreasing extrapolated value with decreasing pion mass.

\begin{figure}[thb!]
\includegraphics[scale=0.5]{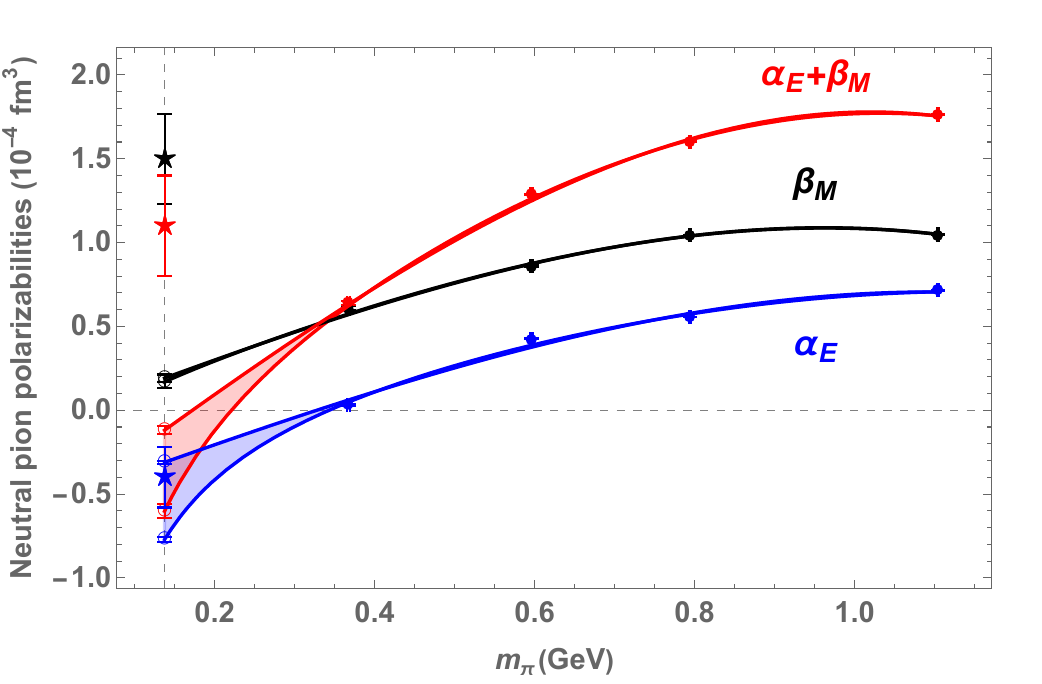}
\caption{
Chiral extrapolation of neutral pion  polarizabilities. The stars at the physical point are from ChPT given in Eq.\eqref{eq:pizero}.
The spread is from two different forms described in the text. 
}
\label{fig:chiral}
\end{figure}
\begin{table*}[htb!]
\caption{Summary of results as a function of pion mass. The polarizabilities are given in standard units of  $10^{-4}\;\text{fm}^3$. 
The upper (lower) extrapolated value at the physical point corresponds to polynomial ($1/m_\pi$) form, respectively.
}
\label{tab:final}
\begin{tabular}{c}
$      
\renewcommand{\arraystretch}{1.2}
\begin{array}{c|cccc|c}
\hline
  & \text{$\kappa $=0.1520} & \text{$\kappa $=0.1543} & \text{$\kappa $=0.1555} & \text{$\kappa $=0.1565} & \text{physical point} \\
\hline
 m_{\pi }\text{(MeV)} & 1104.7\pm 1.2 & 795.0\pm 1.1 & 596.8\pm 1.4 & 367.7\pm 2.2 & 138 \\
\hline
 \alpha _E & \text{0.716(3)} & \text{0.556(2)} & \text{0.425(3)} & \text{0.030(4)} & \text{-0.31(1)} \\
 \text{} & \text{} & \text{} & \text{} & \text{} & \text{-0.77(2)} \\
 \beta _M & \text{1.048(2)} & \text{1.045(3)} & \text{0.861(4)} & \text{0.61(1)} & \text{0.19(2)} \\
 \text{} & \text{} & \text{} & \text{} & \text{} & \text{0.17(4)} \\
 \alpha _E+\beta _M & \text{1.765(4)} & \text{1.601(3)} & \text{1.286(5)} & \text{0.64(1)} & \text{-0.12(2)} \\
 \text{} & \text{} & \text{} & \text{} & \text{} & \text{-0.60(6)} \\
 \hline
\end{array}
$   
\end{tabular}
\end{table*}
\begin{figure}[thb]
\includegraphics[scale=0.5]{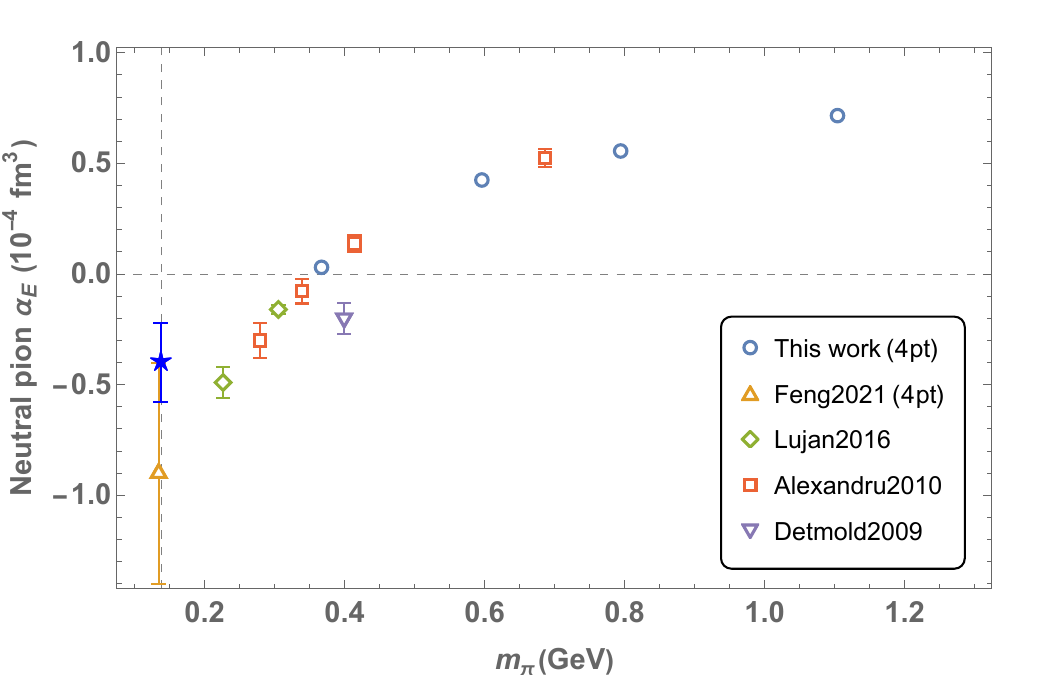}
\caption{
Neutral pion  electric polarizability from background field method and four-point function method (labeled by 4pt). The star at the physical point is from ChPT.
}
\label{fig:alpha}
\end{figure}
\begin{figure}[t!]
\includegraphics[scale=0.5]{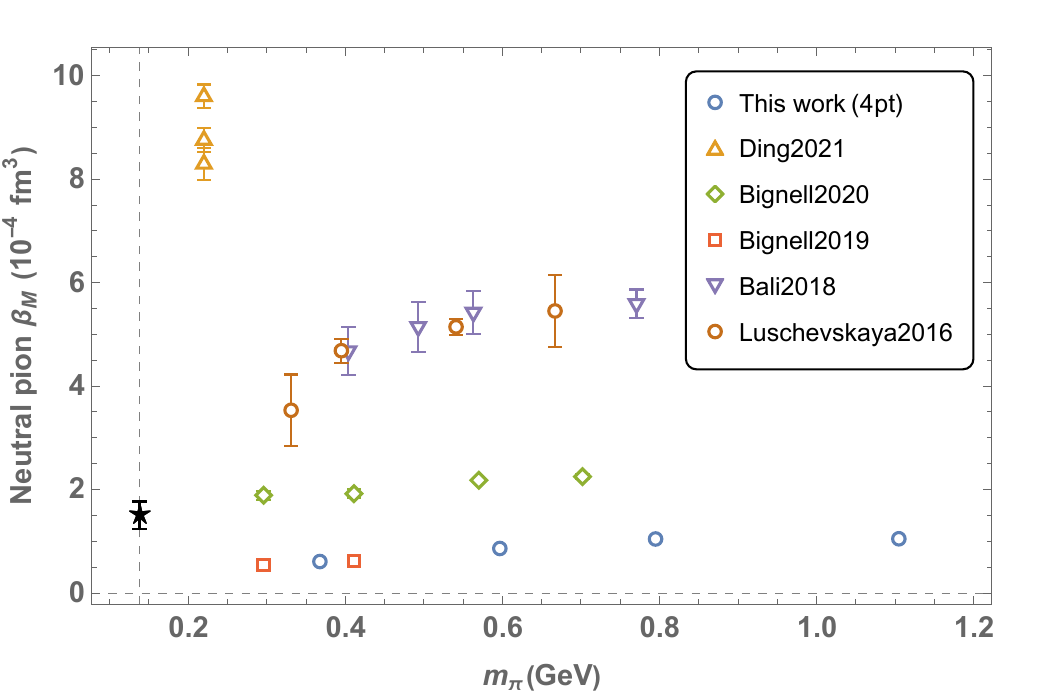}
\caption{
Neutral pion  magnetic polarizability from various lattice QCD calculations. The star at the physical point is from ChPT.
}
\label{fig:beta}
\end{figure}

Finally, we take the extrapolated values and plot them as a function of pion mass in Fig.~\ref{fig:chiral}.
We perform a chiral extrapolation 
to the physical point using  two different forms. One is a polynomial form $a_0+a_1 m_\pi+a_3 m_\pi^3$.
The other is $a_0/m_\pi+a_1 m_\pi+a_3 m_\pi^3$  with a leading $1/m_\pi$ term to account for possible divergencies.
The spread between the two different forms can be regarded a systematic uncertainty.
Although the uncertainty from each fit is comparable, the spread is much smaller for $\beta_M$ than for $\alpha_E$, indicating a mild dependence on the $1/m_\pi$ term for $\beta_M$.
The extrapolation leads to a sign change  for $\alpha_E$ and the extrapolated value is consistent with that from ChPT.
On the other hand the extrapolation for $\beta_M$ leads to a small but positive value that is significantly smaller than that from ChPT.  As a result, the sum of electric and magnetic polarizabilities $\alpha_E+\beta_M$ also tuns negative at the physical point, also at variance with ChPT.
The numbers are summarized in Table~\ref{tab:final}.

In  Fig.~\ref{fig:alpha}, we compare $\alpha_E$ from various lattice calculations in the background field method~\cite{Lujan:2016ffj,Alexandru:2010dx,Detmold:2009dx} and 
the four-point function method.
In addition to our result in this work, there exists another study using four-point function in position space and physical pion mass~\cite{Feng:2022rkr}. By and large, the results are consistent with each other and with ChPT.
So far both methods neglect disconnected contributions. Furthermore, they are electro-quenched in the sea quarks.

 In contrast, the situation for $\beta_M$  is rather different.
 In Fig.~\ref{fig:beta}, we show calculations from the background field method~\cite{Ding2021,Bali:2017ian,Bignell_2020,Bignell_2019,LUSCHEVSKAYA2016393,Luschevskaya:2016epp}, and the sole result in the four-point function method from this work.
We see large disagreements within the background field method, except 
Ref.~\cite{Bali:2017ian,LUSCHEVSKAYA2016393} which agree with each other.  
Notably,  the four-point function result agrees with Ref.~\cite{Bignell_2019} %which uses a standard analysis, 
but disagrees with Ref.~\cite{Bignell_2020} which is an improved analysis of Ref.~\cite{Bignell_2019} based on eigenmode projection. 
%Ref.~\cite{Bali:2017ian,LUSCHEVSKAYA2016393,Bignell_2020} have a smooth approach to the ChPT result; others do not.
The situation calls for more studies in order to understand the physics mechanisms.
A step in this direction is taken in Ref.~\cite{He2020} (also recognized in Ref.~\cite{Hu_2008}) in which a chiral extrapolation of the results in Ref.~\cite{Bignell_2020} is performed using partially-quenched ChPT. In this approach, the neutral pion is dressed by pion cloud at various orders. It identifies the equivalent terms missing in current lattice QCD simulations and estimates the correction to be as large as 1.06 in the standard units.

We would like to mention that a potential source of systematic uncertainty among different calculations in Fig.~\ref{fig:alpha} and Fig.~\ref{fig:beta} is discretization errors in the fermion action. Some of the discrepancies may be due to O(a) errors in the Wilson action and not at all related to difficulties encountered in the polarizability analysis methods.

In the four-point function formalism, we can decompose the polarizabilities into quark components.
Since the formulas for $\pi^0$  in Eq.\eqref{eq:alpha} and Eq.\eqref{eq:beta} are proportional 
to $Q_{44}$ or $Q_{11}$, the relations found in the Wick contractions in the appendix directly translate to polarizabilities, 
\beq
\alpha_E = \alpha^{(CI)}_{uu}+ \alpha^{(DI)}_{uu} +  \alpha^{(CI)}_{dd}+ \alpha^{(DI)}_{dd} + \alpha^{(DI)}_{ud},
\label{eq:qa}
\eeq
\beq
\beta_M = \beta^{(CI)}_{uu}+ \beta^{(DI)}_{uu} +  \beta^{(CI)}_{dd}+ \beta^{(DI)}_{dd} + \beta^{(DI)}_{ud}.
\label{eq:qb}
\eeq
The diagonal uu and dd terms have both connected and disconnected contributions, whereas the ud cross term has only disconnected contributions (it is also absent of diagram D in Fig.~\ref{fig:4pt}). Furthermore, there is an exact relation between the diagonal terms, 
\beq 
\alpha^{(CI)}_{uu} = 4 \alpha^{(CI)}_{dd} \text{ and }  \alpha^{(DI)}_{uu} = 4 \alpha^{(DI)}_{dd},
\label{eq:qa2}
\eeq
\beq 
\beta^{(CI)}_{uu} = 4 \beta^{(CI)}_{dd} \text{ and }  \beta^{(DI)}_{uu} = 4 \beta^{(DI)}_{dd}.
\label{eq:qb2}
\eeq
We emphasize that the relations in  Eq.\eqref{eq:qa} to  Eq.\eqref{eq:qb2} are specific to a neutral pion. For a charged pion, a different decomposition into charge radius, connected, and disconnected contributions exists~\cite{Lee:2023rmz,Lee:2023lnx}. 
For $\beta_M$, we evaluated in this work the connected contributions $ \beta^{(CI)}_{uu}+ \beta^{(CI)}_{dd}=5 \beta^{(CI)}_{dd} = 0.18(2)$ (taking the average of the two values from Table~\ref{tab:final}).
If we regard the value $\beta_M=1.50(27)$ from ChPT in Eq.\eqref{eq:pizero} as the full QCD result, then the difference between ChPT and our result implies a fairly large contribution from the disconnected diagrams $5\beta^{(DI)}_{dd} + \beta^{(DI)}_{ud}  = 1.32(27)$ for $\pi^0$ magnetic polarizability. This compares well with the 1.06 estimated in Ref.~\cite{He2020}. 
%It provides more impetus to evaluate the disconnected diagrams in lattice QCD.

%%%%%%%%%%%%%%%%%%%%%%%%%%%%%%%%%%%%%%%%%%%%%%%%%%%%%%%%%%%%%%%%%%%%
\section{Conclusion}
\label{sec:con}
%%%%%%%%%%%%%%%%%%%%%%%%%%%%%%%%%%%%%%%%%%%%%%%%%%%%%%%%%%%%%%%%%%%%

A neutral pion's electromagnetic polarizabilities offer a unique opportunity to test the QCD-based methods employed to extract them. 
This is mainly due to the different operator structure at the quark level: $\bar{u}\gamma_5 u-\bar{d} \gamma_5 d$ for $\pi^0$  versus $\bar{d}\gamma_5 u$ for $\pi^+$. The former has self-contracting disconnected loops (see the Appendix), while the latter does not. This is true either in two-point or four-point functions.

In this work, we derived new formulas in Eq.\eqref{eq:alpha} and Eq.\eqref{eq:beta} for a neutral pion in the four-point function formalism.
We applied the formulas in a proof-of-concept lattice simulation using the same parameters as for a charged pion~\cite{Lee:2023rmz,Lee:2023lnx}. 
The results for $\alpha_E$ as summarized in Fig.~\ref{fig:alpha} are largely consistent with existing calculations and with ChPT. 
The results for $\beta_M$ as summarized in Fig.~\ref{fig:beta}, on the other hand, are widely inconsistent.

The situation puts a spotlight on the disconnected contributions in neutral pion magnetic polarizability.
Our result from the four-point function method hints a potentially large contribution from the disconnected diagrams. It is also supported by a ChPT-based estimate~\cite{He2020}. 
Our argument  is based on a straightforward  decomposition of the polarizabilities in the four-point function formalism.
Due to the absence of elastic contributions for $\pi^0$, the polarizabilities in Eq.\eqref{eq:alpha} and Eq.\eqref{eq:beta} are proportional to the four-point functions $Q_{44}$ and $Q_{11}$ (albeit under time integrals). Since the four-point functions can be decomposed into quark components of various types (uu, dd, ud) according to their charge factors, the relations in the Appendix translate directly to  polarizabilities as given in Eq.\eqref{eq:qa} to Eq.\eqref{eq:qb2} in which connected and disconnected contributions can be  further separated.  Consequently, one can examine the terms one by one to see their impact on the polarizabilities. In contrast, such contributions are indirectly present via exponential or more complicated functions in the two-point functions used in the background field method. 

Looking forward, the most important issue is to directly simulate the disconnected diagrams on the lattice.
In the meantime, some systematic effects in the current simulation should be addressed, such as $O(a)$ scaling violations in Wilson fermions, and the quenched approximation in the gauge ensembles.
Work is under way to use the $O(a)$-improved two-flavor nhyp-clover ensembles~\cite{Niyazi:2020erg,Brett_2021} to repeat the analysis for both neutral and charged pions. The six dynamical ensembles described in Ref.~\cite{Brett_2021} with elongated geometries also afford the opportunity to study finite-volume effects as well as to reach smaller momentum and pion mass.

%\vspace*{5mm}
\begin{acknowledgments}
This work was supported in part by U.S. Department of Energy under  Grant~No.~DE-FG02-95ER40907 (FL, AA) and UK Research and Innovation grant {MR/S015418/1} (CC). 
WW would like to acknowledge support from the Baylor College of Arts and Sciences SRA program.
Computing resources at DOE-sponsored NERSC and NSF-sponsored TACC were used.
\end{acknowledgments}

%%%%%%%%%%%%%%%%%%%%%%%%%%%%
%\clearpage
\bibliography{xpizero}

\begin{thebibliography}{30}%
\makeatletter
\providecommand \@ifxundefined [1]{%
 \@ifx{#1\undefined}
}%
\providecommand \@ifnum [1]{%
 \ifnum #1\expandafter \@firstoftwo
 \else \expandafter \@secondoftwo
 \fi
}%
\providecommand \@ifx [1]{%
 \ifx #1\expandafter \@firstoftwo
 \else \expandafter \@secondoftwo
 \fi
}%
\providecommand \natexlab [1]{#1}%
\providecommand \enquote  [1]{``#1''}%
\providecommand \bibnamefont  [1]{#1}%
\providecommand \bibfnamefont [1]{#1}%
\providecommand \citenamefont [1]{#1}%
\providecommand \href@noop [0]{\@secondoftwo}%
\providecommand \href [0]{\begingroup \@sanitize@url \@href}%
\providecommand \@href[1]{\@@startlink{#1}\@@href}%
\providecommand \@@href[1]{\endgroup#1\@@endlink}%
\providecommand \@sanitize@url [0]{\catcode `\\12\catcode `\$12\catcode
  `\&12\catcode `\#12\catcode `\^12\catcode `\_12\catcode `\%12\relax}%
\providecommand \@@startlink[1]{}%
\providecommand \@@endlink[0]{}%
\providecommand \url  [0]{\begingroup\@sanitize@url \@url }%
\providecommand \@url [1]{\endgroup\@href {#1}{\urlprefix }}%
\providecommand \urlprefix  [0]{URL }%
\providecommand \Eprint [0]{\href }%
\providecommand \doibase [0]{http://dx.doi.org/}%
\providecommand \selectlanguage [0]{\@gobble}%
\providecommand \bibinfo  [0]{\@secondoftwo}%
\providecommand \bibfield  [0]{\@secondoftwo}%
\providecommand \translation [1]{[#1]}%
\providecommand \BibitemOpen [0]{}%
\providecommand \bibitemStop [0]{}%
\providecommand \bibitemNoStop [0]{.\EOS\space}%
\providecommand \EOS [0]{\spacefactor3000\relax}%
\providecommand \BibitemShut  [1]{\csname bibitem#1\endcsname}%
\let\auto@bib@innerbib\@empty
%</preamble>
\bibitem [{\citenamefont {Endrodi}(2024)}]{endrodi2024}%
  \BibitemOpen
  \bibfield  {author} {\bibinfo {author} {\bibfnamefont {Gergely}\ \bibnamefont
  {Endrodi}},\ }\href {https://arxiv.org/abs/2406.19780} {\enquote {\bibinfo
  {title} {Qcd with background electromagnetic fields on the lattice: a
  review},}\ } (\bibinfo {year} {2024}),\ \Eprint
  {http://arxiv.org/abs/2406.19780} {arXiv:2406.19780 [hep-lat]} \BibitemShut
  {NoStop}%
\bibitem [{\citenamefont {Detmold}\ \emph {et~al.}(2009)\citenamefont
  {Detmold}, \citenamefont {Tiburzi},\ and\ \citenamefont
  {Walker-Loud}}]{Detmold:2009dx}%
  \BibitemOpen
  \bibfield  {author} {\bibinfo {author} {\bibfnamefont {William}\ \bibnamefont
  {Detmold}}, \bibinfo {author} {\bibfnamefont {Brian~C.}\ \bibnamefont
  {Tiburzi}}, \ and\ \bibinfo {author} {\bibfnamefont {Andre}\ \bibnamefont
  {Walker-Loud}},\ }\bibfield  {title} {\enquote {\bibinfo {title} {{Extracting
  Electric Polarizabilities from Lattice QCD}},}\ }\href {\doibase
  10.1103/PhysRevD.79.094505} {\bibfield  {journal} {\bibinfo  {journal} {Phys.
  Rev. D}\ }\textbf {\bibinfo {volume} {79}},\ \bibinfo {pages} {094505}
  (\bibinfo {year} {2009})},\ \Eprint {http://arxiv.org/abs/0904.1586}
  {arXiv:0904.1586 [hep-lat]} \BibitemShut {NoStop}%
\bibitem [{\citenamefont {Niyazi}\ \emph {et~al.}(2021)\citenamefont {Niyazi},
  \citenamefont {Alexandru}, \citenamefont {Lee},\ and\ \citenamefont
  {Lujan}}]{niyazi2021charged}%
  \BibitemOpen
  \bibfield  {author} {\bibinfo {author} {\bibfnamefont {Hossein}\ \bibnamefont
  {Niyazi}}, \bibinfo {author} {\bibfnamefont {Andrei}\ \bibnamefont
  {Alexandru}}, \bibinfo {author} {\bibfnamefont {Frank~X.}\ \bibnamefont
  {Lee}}, \ and\ \bibinfo {author} {\bibfnamefont {Michael}\ \bibnamefont
  {Lujan}},\ }\href@noop {} {\enquote {\bibinfo {title} {Charged pion electric
  polarizability from lattice qcd},}\ } (\bibinfo {year} {2021}),\ \Eprint
  {http://arxiv.org/abs/2105.06906} {arXiv:2105.06906 [hep-lat]} \BibitemShut
  {NoStop}%
\bibitem [{\citenamefont {Bignell}\ \emph {et~al.}(2018)\citenamefont
  {Bignell}, \citenamefont {Hall}, \citenamefont {Kamleh}, \citenamefont
  {Leinweber},\ and\ \citenamefont {Burkardt}}]{Bignell_2018}%
  \BibitemOpen
  \bibfield  {author} {\bibinfo {author} {\bibfnamefont {Ryan}\ \bibnamefont
  {Bignell}}, \bibinfo {author} {\bibfnamefont {Jonathan}\ \bibnamefont
  {Hall}}, \bibinfo {author} {\bibfnamefont {Waseem}\ \bibnamefont {Kamleh}},
  \bibinfo {author} {\bibfnamefont {Derek}\ \bibnamefont {Leinweber}}, \ and\
  \bibinfo {author} {\bibfnamefont {Matthias}\ \bibnamefont {Burkardt}},\
  }\bibfield  {title} {\enquote {\bibinfo {title} {Neutron magnetic
  polarizability with landau mode operators},}\ }\href {\doibase
  10.1103/physrevd.98.034504} {\bibfield  {journal} {\bibinfo  {journal}
  {Physical Review D}\ }\textbf {\bibinfo {volume} {98}} (\bibinfo {year}
  {2018}),\ 10.1103/physrevd.98.034504}\BibitemShut {NoStop}%
\bibitem [{\citenamefont {Bignell}\ \emph
  {et~al.}(2020{\natexlab{a}})\citenamefont {Bignell}, \citenamefont {Kamleh},\
  and\ \citenamefont {Leinweber}}]{Bignell_2020}%
  \BibitemOpen
  \bibfield  {author} {\bibinfo {author} {\bibfnamefont {Ryan}\ \bibnamefont
  {Bignell}}, \bibinfo {author} {\bibfnamefont {Waseem}\ \bibnamefont
  {Kamleh}}, \ and\ \bibinfo {author} {\bibfnamefont {Derek}\ \bibnamefont
  {Leinweber}},\ }\bibfield  {title} {\enquote {\bibinfo {title} {Pion magnetic
  polarisability using the background field method},}\ }\href {\doibase
  10.1016/j.physletb.2020.135853} {\bibfield  {journal} {\bibinfo  {journal}
  {Physics Letters B}\ }\textbf {\bibinfo {volume} {811}},\ \bibinfo {pages}
  {135853} (\bibinfo {year} {2020}{\natexlab{a}})}\BibitemShut {NoStop}%
\bibitem [{\citenamefont {Bignell}\ \emph
  {et~al.}(2020{\natexlab{b}})\citenamefont {Bignell}, \citenamefont {Kamleh},\
  and\ \citenamefont {Leinweber}}]{Bignell:2020xkf}%
  \BibitemOpen
  \bibfield  {author} {\bibinfo {author} {\bibfnamefont {Ryan}\ \bibnamefont
  {Bignell}}, \bibinfo {author} {\bibfnamefont {Waseem}\ \bibnamefont
  {Kamleh}}, \ and\ \bibinfo {author} {\bibfnamefont {Derek}\ \bibnamefont
  {Leinweber}},\ }\bibfield  {title} {\enquote {\bibinfo {title} {{Magnetic
  polarizability of the nucleon using a Laplacian mode projection}},}\ }\href
  {\doibase 10.1103/PhysRevD.101.094502} {\bibfield  {journal} {\bibinfo
  {journal} {Phys. Rev. D}\ }\textbf {\bibinfo {volume} {101}},\ \bibinfo
  {pages} {094502} (\bibinfo {year} {2020}{\natexlab{b}})},\ \Eprint
  {http://arxiv.org/abs/2002.07915} {arXiv:2002.07915 [hep-lat]} \BibitemShut
  {NoStop}%
\bibitem [{\citenamefont {He}\ \emph {et~al.}(2021)\citenamefont {He},
  \citenamefont {Leinweber}, \citenamefont {Thomas},\ and\ \citenamefont
  {Wang}}]{He:2021eha}%
  \BibitemOpen
  \bibfield  {author} {\bibinfo {author} {\bibfnamefont {Fangcheng}\
  \bibnamefont {He}}, \bibinfo {author} {\bibfnamefont {Derek~B.}\ \bibnamefont
  {Leinweber}}, \bibinfo {author} {\bibfnamefont {Anthony~W.}\ \bibnamefont
  {Thomas}}, \ and\ \bibinfo {author} {\bibfnamefont {Ping}\ \bibnamefont
  {Wang}},\ }\bibfield  {title} {\enquote {\bibinfo {title} {{Chiral
  extrapolation of the charged-pion magnetic polarizability with Pad\'e
  approximant}},}\ }\href@noop {} {\  (\bibinfo {year} {2021})},\ \Eprint
  {http://arxiv.org/abs/2104.09963} {arXiv:2104.09963 [nucl-th]} \BibitemShut
  {NoStop}%
\bibitem [{\citenamefont {Burkardt}\ \emph {et~al.}(1995)\citenamefont
  {Burkardt}, \citenamefont {Grandy},\ and\ \citenamefont
  {Negele}}]{BURKARDT1995441}%
  \BibitemOpen
  \bibfield  {author} {\bibinfo {author} {\bibfnamefont {M.}~\bibnamefont
  {Burkardt}}, \bibinfo {author} {\bibfnamefont {J.M.}\ \bibnamefont {Grandy}},
  \ and\ \bibinfo {author} {\bibfnamefont {J.W.}\ \bibnamefont {Negele}},\
  }\bibfield  {title} {\enquote {\bibinfo {title} {Calculation and
  interpretation of hadron correlation functions in lattice qcd},}\ }\href
  {\doibase https://doi.org/10.1006/aphy.1995.1026} {\bibfield  {journal}
  {\bibinfo  {journal} {Annals of Physics}\ }\textbf {\bibinfo {volume}
  {238}},\ \bibinfo {pages} {441--472} (\bibinfo {year} {1995})}\BibitemShut
  {NoStop}%
\bibitem [{\citenamefont {Andersen}\ and\ \citenamefont
  {Wilcox}(1997)}]{Andersen:1996qb}%
  \BibitemOpen
  \bibfield  {author} {\bibinfo {author} {\bibfnamefont {William}\ \bibnamefont
  {Andersen}}\ and\ \bibinfo {author} {\bibfnamefont {Walter}\ \bibnamefont
  {Wilcox}},\ }\bibfield  {title} {\enquote {\bibinfo {title} {{Lattice charge
  overlap. 1. Elastic limit of pi and rho mesons}},}\ }\href {\doibase
  10.1006/aphy.1996.5648} {\bibfield  {journal} {\bibinfo  {journal} {Annals
  Phys.}\ }\textbf {\bibinfo {volume} {255}},\ \bibinfo {pages} {34--59}
  (\bibinfo {year} {1997})},\ \Eprint {http://arxiv.org/abs/hep-lat/9502015}
  {arXiv:hep-lat/9502015} \BibitemShut {NoStop}%
\bibitem [{\citenamefont {Wilcox}(1997)}]{Wilcox:1996vx}%
  \BibitemOpen
  \bibfield  {author} {\bibinfo {author} {\bibfnamefont {Walter}\ \bibnamefont
  {Wilcox}},\ }\bibfield  {title} {\enquote {\bibinfo {title} {{Lattice charge
  overlap. 2: Aspects of charged pion polarizability}},}\ }\href {\doibase
  10.1006/aphy.1996.5649} {\bibfield  {journal} {\bibinfo  {journal} {Annals
  Phys.}\ }\textbf {\bibinfo {volume} {255}},\ \bibinfo {pages} {60--74}
  (\bibinfo {year} {1997})},\ \Eprint {http://arxiv.org/abs/hep-lat/9606019}
  {arXiv:hep-lat/9606019} \BibitemShut {NoStop}%
\bibitem [{\citenamefont {Engelhardt}(2007)}]{Engelhardt:2007ub}%
  \BibitemOpen
  \bibfield  {author} {\bibinfo {author} {\bibfnamefont {Michael}\ \bibnamefont
  {Engelhardt}},\ }\bibfield  {title} {\enquote {\bibinfo {title} {{Neutron
  electric polarizability from unquenched lattice QCD using the background
  field approach}},}\ }\href {\doibase 10.1103/PhysRevD.76.114502} {\bibfield
  {journal} {\bibinfo  {journal} {Phys. Rev. D}\ }\textbf {\bibinfo {volume}
  {76}},\ \bibinfo {pages} {114502} (\bibinfo {year} {2007})},\ \Eprint
  {http://arxiv.org/abs/0706.3919} {arXiv:0706.3919 [hep-lat]} \BibitemShut
  {NoStop}%
\bibitem [{\citenamefont {Wilcox}\ and\ \citenamefont
  {Lee}(2021)}]{Wilcox:2021rtt}%
  \BibitemOpen
  \bibfield  {author} {\bibinfo {author} {\bibfnamefont {Walter}\ \bibnamefont
  {Wilcox}}\ and\ \bibinfo {author} {\bibfnamefont {Frank~X.}\ \bibnamefont
  {Lee}},\ }\bibfield  {title} {\enquote {\bibinfo {title} {{Towards charged
  hadron polarizabilities from four-point functions in lattice QCD}},}\ }\href
  {\doibase 10.1103/PhysRevD.104.034506} {\bibfield  {journal} {\bibinfo
  {journal} {Phys. Rev. D}\ }\textbf {\bibinfo {volume} {104}},\ \bibinfo
  {pages} {034506} (\bibinfo {year} {2021})},\ \Eprint
  {http://arxiv.org/abs/2106.02557} {arXiv:2106.02557 [hep-lat]} \BibitemShut
  {NoStop}%
\bibitem [{\citenamefont {Lee}\ \emph {et~al.}(2023{\natexlab{a}})\citenamefont
  {Lee}, \citenamefont {Alexandru}, \citenamefont {Culver},\ and\ \citenamefont
  {Wilcox}}]{Lee:2023rmz}%
  \BibitemOpen
  \bibfield  {author} {\bibinfo {author} {\bibfnamefont {Frank~X.}\
  \bibnamefont {Lee}}, \bibinfo {author} {\bibfnamefont {Andrei}\ \bibnamefont
  {Alexandru}}, \bibinfo {author} {\bibfnamefont {Chris}\ \bibnamefont
  {Culver}}, \ and\ \bibinfo {author} {\bibfnamefont {Walter}\ \bibnamefont
  {Wilcox}},\ }\bibfield  {title} {\enquote {\bibinfo {title} {{Charged pion
  electric polarizability from four-point functions in lattice QCD}},}\
  }\href@noop {} {\  (\bibinfo {year} {2023}{\natexlab{a}})},\ \Eprint
  {http://arxiv.org/abs/2301.05200} {arXiv:2301.05200 [hep-lat]} \BibitemShut
  {NoStop}%
\bibitem [{\citenamefont {Lee}\ \emph {et~al.}(2023{\natexlab{b}})\citenamefont
  {Lee}, \citenamefont {Wilcox}, \citenamefont {Alexandru},\ and\ \citenamefont
  {Culver}}]{Lee:2023lnx}%
  \BibitemOpen
  \bibfield  {author} {\bibinfo {author} {\bibfnamefont {Frank~X.}\
  \bibnamefont {Lee}}, \bibinfo {author} {\bibfnamefont {Walter}\ \bibnamefont
  {Wilcox}}, \bibinfo {author} {\bibfnamefont {Andrei}\ \bibnamefont
  {Alexandru}}, \ and\ \bibinfo {author} {\bibfnamefont {Chris}\ \bibnamefont
  {Culver}},\ }\bibfield  {title} {\enquote {\bibinfo {title} {{Magnetic
  polarizability of a charged pion from four-point functions in lattice
  QCD}},}\ }\href {\doibase 10.1103/PhysRevD.108.054510} {\bibfield  {journal}
  {\bibinfo  {journal} {Phys. Rev. D}\ }\textbf {\bibinfo {volume} {108}},\
  \bibinfo {pages} {054510} (\bibinfo {year} {2023}{\natexlab{b}})},\ \Eprint
  {http://arxiv.org/abs/2307.08620} {arXiv:2307.08620 [hep-lat]} \BibitemShut
  {NoStop}%
\bibitem [{\citenamefont {Feng}\ \emph {et~al.}(2022)\citenamefont {Feng},
  \citenamefont {Izubuchi}, \citenamefont {Jin},\ and\ \citenamefont
  {Golterman}}]{Feng:2022rkr}%
  \BibitemOpen
  \bibfield  {author} {\bibinfo {author} {\bibfnamefont {Xu}~\bibnamefont
  {Feng}}, \bibinfo {author} {\bibfnamefont {Taku}\ \bibnamefont {Izubuchi}},
  \bibinfo {author} {\bibfnamefont {Luchang}\ \bibnamefont {Jin}}, \ and\
  \bibinfo {author} {\bibfnamefont {Maarten}\ \bibnamefont {Golterman}},\
  }\bibfield  {title} {\enquote {\bibinfo {title} {{Pion electric
  polarizabilities from lattice QCD}},}\ }\href {\doibase 10.22323/1.396.0362}
  {\bibfield  {journal} {\bibinfo  {journal} {PoS}\ }\textbf {\bibinfo {volume}
  {LATTICE2021}},\ \bibinfo {pages} {362} (\bibinfo {year} {2022})},\ \Eprint
  {http://arxiv.org/abs/2201.01396} {arXiv:2201.01396 [hep-lat]} \BibitemShut
  {NoStop}%
\bibitem [{\citenamefont {Wang}\ \emph {et~al.}(2024)\citenamefont {Wang},
  \citenamefont {Zhang}, \citenamefont {Cao}, \citenamefont {Fan},
  \citenamefont {Feng}, \citenamefont {Gao}, \citenamefont {Jin},\ and\
  \citenamefont {Liu}}]{wang2024}%
  \BibitemOpen
  \bibfield  {author} {\bibinfo {author} {\bibfnamefont {Xuan-He}\ \bibnamefont
  {Wang}}, \bibinfo {author} {\bibfnamefont {Zhao-Long}\ \bibnamefont {Zhang}},
  \bibinfo {author} {\bibfnamefont {Xiong-Hui}\ \bibnamefont {Cao}}, \bibinfo
  {author} {\bibfnamefont {Cong-Ling}\ \bibnamefont {Fan}}, \bibinfo {author}
  {\bibfnamefont {Xu}~\bibnamefont {Feng}}, \bibinfo {author} {\bibfnamefont
  {Yu-Sheng}\ \bibnamefont {Gao}}, \bibinfo {author} {\bibfnamefont {Lu-Chang}\
  \bibnamefont {Jin}}, \ and\ \bibinfo {author} {\bibfnamefont {Chuan}\
  \bibnamefont {Liu}},\ }\href {https://arxiv.org/abs/2310.01168} {\enquote
  {\bibinfo {title} {Nucleon electric polarizabilities and nucleon-pion
  scattering at physical pion mass},}\ } (\bibinfo {year} {2024}),\ \Eprint
  {http://arxiv.org/abs/2310.01168} {arXiv:2310.01168 [hep-lat]} \BibitemShut
  {NoStop}%
\bibitem [{\citenamefont {B{\"u}rgi}(1996)}]{BURGI1996392}%
  \BibitemOpen
  \bibfield  {author} {\bibinfo {author} {\bibfnamefont {U}~\bibnamefont
  {B{\"u}rgi}},\ }\bibfield  {title} {\enquote {\bibinfo {title} {Pion
  polarizabilities and charged pion-pair production to two loops},}\ }\href
  {\doibase https://doi.org/10.1016/0550-3213(96)00454-3} {\bibfield  {journal}
  {\bibinfo  {journal} {Nuclear Physics B}\ }\textbf {\bibinfo {volume}
  {479}},\ \bibinfo {pages} {392--426} (\bibinfo {year} {1996})}\BibitemShut
  {NoStop}%
\bibitem [{\citenamefont {Gasser}\ \emph {et~al.}(2006)\citenamefont {Gasser},
  \citenamefont {Ivanov},\ and\ \citenamefont {Sainio}}]{Gasser_2006}%
  \BibitemOpen
  \bibfield  {author} {\bibinfo {author} {\bibfnamefont {J.}~\bibnamefont
  {Gasser}}, \bibinfo {author} {\bibfnamefont {M.A.}\ \bibnamefont {Ivanov}}, \
  and\ \bibinfo {author} {\bibfnamefont {M.E.}\ \bibnamefont {Sainio}},\
  }\bibfield  {title} {\enquote {\bibinfo {title} {Revisiting gamma + gamma to
  pi+ and pi- at low energies},}\ }\href {\doibase
  10.1016/j.nuclphysb.2006.03.022} {\bibfield  {journal} {\bibinfo  {journal}
  {Nuclear Physics B}\ }\textbf {\bibinfo {volume} {745}},\ \bibinfo {pages}
  {84--108} (\bibinfo {year} {2006})}\BibitemShut {NoStop}%
\bibitem [{\citenamefont {Moinester}\ and\ \citenamefont
  {Scherer}(2019)}]{Moinester:2019sew}%
  \BibitemOpen
  \bibfield  {author} {\bibinfo {author} {\bibfnamefont {Murray}\ \bibnamefont
  {Moinester}}\ and\ \bibinfo {author} {\bibfnamefont {Stefan}\ \bibnamefont
  {Scherer}},\ }\bibfield  {title} {\enquote {\bibinfo {title} {{Compton
  Scattering off Pions and Electromagnetic Polarizabilities}},}\ }\href
  {\doibase 10.1142/S0217751X19300084} {\bibfield  {journal} {\bibinfo
  {journal} {Int. J. Mod. Phys. A}\ }\textbf {\bibinfo {volume} {34}},\
  \bibinfo {pages} {1930008} (\bibinfo {year} {2019})},\ \Eprint
  {http://arxiv.org/abs/1905.05640} {arXiv:1905.05640 [hep-ph]} \BibitemShut
  {NoStop}%
\bibitem [{\citenamefont {He}\ \emph {et~al.}(2020)\citenamefont {He},
  \citenamefont {Leinweber}, \citenamefont {Thomas},\ and\ \citenamefont
  {Wang}}]{He2020}%
  \BibitemOpen
  \bibfield  {author} {\bibinfo {author} {\bibfnamefont {Fangcheng}\
  \bibnamefont {He}}, \bibinfo {author} {\bibfnamefont {D.~B.}\ \bibnamefont
  {Leinweber}}, \bibinfo {author} {\bibfnamefont {A.~W.}\ \bibnamefont
  {Thomas}}, \ and\ \bibinfo {author} {\bibfnamefont {P.}~\bibnamefont
  {Wang}},\ }\bibfield  {title} {\enquote {\bibinfo {title} {Chiral
  extrapolation of the magnetic polarizability of the neutral pion},}\ }\href
  {\doibase 10.1103/PhysRevD.102.114509} {\bibfield  {journal} {\bibinfo
  {journal} {Phys. Rev. D}\ }\textbf {\bibinfo {volume} {102}},\ \bibinfo
  {pages} {114509} (\bibinfo {year} {2020})}\BibitemShut {NoStop}%
\bibitem [{\citenamefont {Lujan}\ \emph {et~al.}(2016)\citenamefont {Lujan},
  \citenamefont {Alexandru}, \citenamefont {Freeman},\ and\ \citenamefont
  {Lee}}]{Lujan:2016ffj}%
  \BibitemOpen
  \bibfield  {author} {\bibinfo {author} {\bibfnamefont {M.}~\bibnamefont
  {Lujan}}, \bibinfo {author} {\bibfnamefont {A.}~\bibnamefont {Alexandru}},
  \bibinfo {author} {\bibfnamefont {W.}~\bibnamefont {Freeman}}, \ and\
  \bibinfo {author} {\bibfnamefont {F.~X.}\ \bibnamefont {Lee}},\ }\bibfield
  {title} {\enquote {\bibinfo {title} {{Finite volume effects on the electric
  polarizability of neutral hadrons in lattice QCD}},}\ }\href {\doibase
  10.1103/PhysRevD.94.074506} {\bibfield  {journal} {\bibinfo  {journal} {Phys.
  Rev.}\ }\textbf {\bibinfo {volume} {D94}},\ \bibinfo {pages} {074506}
  (\bibinfo {year} {2016})},\ \Eprint {http://arxiv.org/abs/1606.07928}
  {arXiv:1606.07928 [hep-lat]} \BibitemShut {NoStop}%
%%CITATION = ARXIV:1606.07928;%%
\bibitem [{\citenamefont {Alexandru}\ and\ \citenamefont
  {Lee}(2010)}]{Alexandru:2010dx}%
  \BibitemOpen
  \bibfield  {author} {\bibinfo {author} {\bibfnamefont {Andrei}\ \bibnamefont
  {Alexandru}}\ and\ \bibinfo {author} {\bibfnamefont {Frank}\ \bibnamefont
  {Lee}},\ }\bibfield  {title} {\enquote {\bibinfo {title} {{Hadron electric
  polarizability -- finite volume corrections}},}\ }\href@noop {} {\bibfield
  {journal} {\bibinfo  {journal} {PoS}\ }\textbf {\bibinfo {volume}
  {LATTICE2010}},\ \bibinfo {pages} {131} (\bibinfo {year} {2010})},\ \Eprint
  {http://arxiv.org/abs/1011.6309} {arXiv:1011.6309 [hep-lat]} \BibitemShut
  {NoStop}%
\bibitem [{\citenamefont {Ding}\ \emph {et~al.}(2021)\citenamefont {Ding},
  \citenamefont {Li}, \citenamefont {Tomiya}, \citenamefont {Wang},\ and\
  \citenamefont {Zhang}}]{Ding2021}%
  \BibitemOpen
  \bibfield  {author} {\bibinfo {author} {\bibfnamefont {H.-T.}\ \bibnamefont
  {Ding}}, \bibinfo {author} {\bibfnamefont {S.-T.}\ \bibnamefont {Li}},
  \bibinfo {author} {\bibfnamefont {A.}~\bibnamefont {Tomiya}}, \bibinfo
  {author} {\bibfnamefont {X.-D.}\ \bibnamefont {Wang}}, \ and\ \bibinfo
  {author} {\bibfnamefont {Y.}~\bibnamefont {Zhang}},\ }\bibfield  {title}
  {\enquote {\bibinfo {title} {Chiral properties of ($2+1$)-flavor qcd in
  strong magnetic fields at zero temperature},}\ }\href {\doibase
  10.1103/PhysRevD.104.014505} {\bibfield  {journal} {\bibinfo  {journal}
  {Phys. Rev. D}\ }\textbf {\bibinfo {volume} {104}},\ \bibinfo {pages}
  {014505} (\bibinfo {year} {2021})}\BibitemShut {NoStop}%
\bibitem [{\citenamefont {Bali}\ \emph {et~al.}(2018)\citenamefont {Bali},
  \citenamefont {Brandt}, \citenamefont {Endr\H{o}di},\ and\ \citenamefont
  {Gl\"a\ss{}le}}]{Bali:2017ian}%
  \BibitemOpen
  \bibfield  {author} {\bibinfo {author} {\bibfnamefont {Gunnar~S.}\
  \bibnamefont {Bali}}, \bibinfo {author} {\bibfnamefont {Bastian~B.}\
  \bibnamefont {Brandt}}, \bibinfo {author} {\bibfnamefont {Gergely}\
  \bibnamefont {Endr\H{o}di}}, \ and\ \bibinfo {author} {\bibfnamefont
  {Benjamin}\ \bibnamefont {Gl\"a\ss{}le}},\ }\bibfield  {title} {\enquote
  {\bibinfo {title} {{Meson masses in electromagnetic fields with Wilson
  fermions}},}\ }\href {\doibase 10.1103/PhysRevD.97.034505} {\bibfield
  {journal} {\bibinfo  {journal} {Phys. Rev. D}\ }\textbf {\bibinfo {volume}
  {97}},\ \bibinfo {pages} {034505} (\bibinfo {year} {2018})},\ \Eprint
  {http://arxiv.org/abs/1707.05600} {arXiv:1707.05600 [hep-lat]} \BibitemShut
  {NoStop}%
\bibitem [{\citenamefont {Bignell}\ \emph {et~al.}(2019)\citenamefont
  {Bignell}, \citenamefont {Kamleh},\ and\ \citenamefont
  {Leinweber}}]{Bignell_2019}%
  \BibitemOpen
  \bibfield  {author} {\bibinfo {author} {\bibfnamefont {Ryan}\ \bibnamefont
  {Bignell}}, \bibinfo {author} {\bibfnamefont {Waseem}\ \bibnamefont
  {Kamleh}}, \ and\ \bibinfo {author} {\bibfnamefont {Derek}\ \bibnamefont
  {Leinweber}},\ }\bibfield  {title} {\enquote {\bibinfo {title} {Pion in a
  uniform background magnetic field with clover fermions},}\ }\href {\doibase
  10.1103/physrevd.100.114518} {\bibfield  {journal} {\bibinfo  {journal}
  {Physical Review D}\ }\textbf {\bibinfo {volume} {100}} (\bibinfo {year}
  {2019}),\ 10.1103/physrevd.100.114518}\BibitemShut {NoStop}%
\bibitem [{\citenamefont {Luschevskaya}\ \emph {et~al.}(2016)\citenamefont
  {Luschevskaya}, \citenamefont {Solovjeva},\ and\ \citenamefont
  {Teryaev}}]{LUSCHEVSKAYA2016393}%
  \BibitemOpen
  \bibfield  {author} {\bibinfo {author} {\bibfnamefont {E.V.}\ \bibnamefont
  {Luschevskaya}}, \bibinfo {author} {\bibfnamefont {O.E.}\ \bibnamefont
  {Solovjeva}}, \ and\ \bibinfo {author} {\bibfnamefont {O.V.}\ \bibnamefont
  {Teryaev}},\ }\bibfield  {title} {\enquote {\bibinfo {title} {Magnetic
  polarizability of pion},}\ }\href {\doibase
  https://doi.org/10.1016/j.physletb.2016.08.054} {\bibfield  {journal}
  {\bibinfo  {journal} {Physics Letters B}\ }\textbf {\bibinfo {volume}
  {761}},\ \bibinfo {pages} {393--398} (\bibinfo {year} {2016})}\BibitemShut
  {NoStop}%
\bibitem [{\citenamefont {Luschevskaya}\ \emph {et~al.}(2017)\citenamefont
  {Luschevskaya}, \citenamefont {Solovjeva},\ and\ \citenamefont
  {Teryaev}}]{Luschevskaya:2016epp}%
  \BibitemOpen
  \bibfield  {author} {\bibinfo {author} {\bibfnamefont {E.~V.}\ \bibnamefont
  {Luschevskaya}}, \bibinfo {author} {\bibfnamefont {O.~E.}\ \bibnamefont
  {Solovjeva}}, \ and\ \bibinfo {author} {\bibfnamefont {O.~V.}\ \bibnamefont
  {Teryaev}},\ }\bibfield  {title} {\enquote {\bibinfo {title} {{Determination
  of the properties of vector mesons in external magnetic field by Quenched
  $SU(3)$ Lattice QCD}},}\ }\href {\doibase 10.1007/JHEP09(2017)142} {\bibfield
   {journal} {\bibinfo  {journal} {JHEP}\ }\textbf {\bibinfo {volume} {09}},\
  \bibinfo {pages} {142} (\bibinfo {year} {2017})},\ \Eprint
  {http://arxiv.org/abs/1608.03472} {arXiv:1608.03472 [hep-lat]} \BibitemShut
  {NoStop}%
\bibitem [{\citenamefont {Hu}\ \emph {et~al.}(2008)\citenamefont {Hu},
  \citenamefont {Jiang},\ and\ \citenamefont {Tiburzi}}]{Hu_2008}%
  \BibitemOpen
  \bibfield  {author} {\bibinfo {author} {\bibfnamefont {Jie}\ \bibnamefont
  {Hu}}, \bibinfo {author} {\bibfnamefont {Fu-Jiun}\ \bibnamefont {Jiang}}, \
  and\ \bibinfo {author} {\bibfnamefont {Brian~C.}\ \bibnamefont {Tiburzi}},\
  }\bibfield  {title} {\enquote {\bibinfo {title} {Pion polarizabilities and
  volume effects in lattice qcd},}\ }\href {\doibase
  10.1103/physrevd.77.014502} {\bibfield  {journal} {\bibinfo  {journal}
  {Physical Review D}\ }\textbf {\bibinfo {volume} {77}} (\bibinfo {year}
  {2008}),\ 10.1103/physrevd.77.014502}\BibitemShut {NoStop}%
\bibitem [{\citenamefont {Niyazi}\ \emph {et~al.}(2020)\citenamefont {Niyazi},
  \citenamefont {Alexandru}, \citenamefont {Lee},\ and\ \citenamefont
  {Brett}}]{Niyazi:2020erg}%
  \BibitemOpen
  \bibfield  {author} {\bibinfo {author} {\bibfnamefont {Hossein}\ \bibnamefont
  {Niyazi}}, \bibinfo {author} {\bibfnamefont {Andrei}\ \bibnamefont
  {Alexandru}}, \bibinfo {author} {\bibfnamefont {Frank~X.}\ \bibnamefont
  {Lee}}, \ and\ \bibinfo {author} {\bibfnamefont {Ruair\'\i{}}\ \bibnamefont
  {Brett}},\ }\bibfield  {title} {\enquote {\bibinfo {title} {{Setting the
  scale for nHYP fermions with the L\"uscher-Weisz gauge action}},}\ }\href
  {\doibase 10.1103/PhysRevD.102.094506} {\bibfield  {journal} {\bibinfo
  {journal} {Phys. Rev. D}\ }\textbf {\bibinfo {volume} {102}},\ \bibinfo
  {pages} {094506} (\bibinfo {year} {2020})},\ \Eprint
  {http://arxiv.org/abs/2008.13022} {arXiv:2008.13022 [hep-lat]} \BibitemShut
  {NoStop}%
\bibitem [{\citenamefont {Brett}\ \emph {et~al.}(2021)\citenamefont {Brett},
  \citenamefont {Culver}, \citenamefont {Mai}, \citenamefont {Alexandru},
  \citenamefont {D{\"o}ring},\ and\ \citenamefont {Lee}}]{Brett_2021}%
  \BibitemOpen
  \bibfield  {author} {\bibinfo {author} {\bibfnamefont {Ruair{\'\i}}\
  \bibnamefont {Brett}}, \bibinfo {author} {\bibfnamefont {Chris}\ \bibnamefont
  {Culver}}, \bibinfo {author} {\bibfnamefont {Maxim}\ \bibnamefont {Mai}},
  \bibinfo {author} {\bibfnamefont {Andrei}\ \bibnamefont {Alexandru}},
  \bibinfo {author} {\bibfnamefont {Michael}\ \bibnamefont {D{\"o}ring}}, \
  and\ \bibinfo {author} {\bibfnamefont {Frank~X.}\ \bibnamefont {Lee}},\
  }\bibfield  {title} {\enquote {\bibinfo {title} {Three-body interactions from
  the finite-volume qcd spectrum},}\ }\href {\doibase
  10.1103/physrevd.104.014501} {\bibfield  {journal} {\bibinfo  {journal}
  {Physical Review D}\ }\textbf {\bibinfo {volume} {104}} (\bibinfo {year}
  {2021}),\ 10.1103/physrevd.104.014501}\BibitemShut {NoStop}%
\end{thebibliography}%
%%%%%%%%%%%%%%%%%%%%%%%%%%%%

%%%%%%%%%%%%%%%%%%%%%%%%%%%%%%%%%%%%%%%%%%%%%%%%%%%%%%%%%%%
\begin{widetext}
\appendix
\section{Wick contractions}
\label{sec:wick}
%%%%%%%%%%%%%%%%%%%%%%%%%%%%%%%%%%%%%%%%%%%%%%%%%%%%%%%%%%%
Here we give the  correlation functions in Eq.\eqref{eq:Q11} by contracting out all quark-antiqurk pairs.
%A neutral pion has a different structure from a charged pion in that 
%even at the operator level  the former has self-contracting loops while the latter does not.
We use local current to demonstrate the various parts of the Wick contractions because of its simplicity.  However, a parallel version below regarding the three types exists for conserved current. 
All numerical results in this work are based on conserved current. 
 
  First, we consider the two-point function that serves as normalization to the four-point function in Eq.\eqref{eq:Q11}.
  We will leave out the overall factor $1/\sqrt{2}$  in the interpolating field of Eq.\eqref{eq:op} in both the denominator and numerator.
  Keeping u and d quark labels explicit, we have
 \beqs
d_{1}=\text{tr}\left[S_{u}(t_0, t_3)\gamma_5S_{u}(t_3, t_0)\gamma_5\right]\\
d_{5}=\text{tr}\left[S_{d}(t_0, t_3)\gamma_5S_{d}(t_3, t_0)\gamma_5\right] \\
d_{0}=-\text{tr}\left[S_{u}(t_3, t_3)\gamma_5\right]\text{tr}\left[S_{u}(t_0, t_0)\gamma_5\right]\\
d_{2}=\text{tr}\left[S_{u}(t_3, t_3)\gamma_5\right]\text{tr}\left[S_{d}(t_0, t_0)\gamma_5\right]\\
d_{3}=\text{tr}\left[S_{d}(t_3, t_3)\gamma_5\right]\text{tr}\left[S_{u}(t_0, t_0)\gamma_5\right]\\
d_{4}=-\text{tr}\left[S_{d}(t_3, t_3)\gamma_5\right]\text{tr}\left[S_{d}(t_0, t_0)\gamma_5\right].
\eeqs
 The trace is over spin and color. 
We use a matrix notation that highlights time dependence.
For example,  $S(t_3,t_0)$ denotes a fully-interacting 12x12 quark propagator in spin-color space  from $t_0$ to $t_3$, obtained from an inversion of the quark matrix $M$ with a source ($M x=b$). The spatial sums over $(\bm x_3, \bm x_0)$ are implicit.

We see that in the isospin limit ($\kappa_u=\kappa_d=\kappa$), the disconnected loops cancel, 
leaving only a connected contribution,
\beq
d_{1}+d_{5}=2\,\text{tr}\left[S(t_0, t_3)\gamma_5S(t_3, t_0)\gamma_5\right].
  \label{eq:2pt}
\eeq
This is an important result: in the isospin limit, charged and neutral pions have the same two-point function as normalization. The difference is in the four-point functions. It is worth pointing out that in the background field method, the disconnected loops no longer cancel due to breaking of isospin symmetry by the background field.

Next, we consider the unnormalized four-point functions (denoted by $ \tilde{Q}_{\mu\mu}$ instead of ${Q}_{\mu\mu}$) in the numerator of Eq.\eqref{eq:Q11}, using the local current in Eq.\eqref{eq:j1PC}.
We sort them into diagonal and crossed contributions based on $\bar{u}\gamma_\mu u$ or $\bar{d}\gamma_\mu d$  in the current-current correlations, 
\beqs
   \langle \Omega | \psi_{\pi^0}j^{(PC)}_{\mu} j^{(PC)}_{\mu}\psi_{\pi^0}^\dagger| \Omega\rangle 
%= \langle \Omega | \psi_{\pi^0}\big(q_u \bar{u}\gamma_\mu u + q_d \bar{d}\gamma_\mu d \big)\big(q_u \bar{u}\gamma_\mu u + q_d \bar{d}\gamma_\mu d \big)\psi_{\pi^0}^\dagger| \Omega\rangle   
&= q_u^2 \langle \Omega | \psi_{\pi^0}\big(\bar{u}\gamma_\mu u \big)\big(\bar{u}\gamma_\mu u \big)\psi_{\pi^0}^\dagger| \Omega\rangle   
+q_d^2 \langle \Omega | \psi_{\pi^0}\big( \bar{d}\gamma_\mu d \big)\big( \bar{d}\gamma_\mu d \big)\psi_{\pi^0}^\dagger| \Omega\rangle  \\
&+ q_u q_d\big[ \langle \Omega | \psi_{\pi^0}\big( \bar{u}\gamma_\mu u \big)\big( \bar{d}\gamma_\mu d \big)\psi_{\pi^0}^\dagger| \Omega\rangle   
+\langle \Omega | \psi_{\pi^0}\big( \bar{d}\gamma_\mu d \big)\big( \bar{u}\gamma_\mu u \big)\psi_{\pi^0}^\dagger| \Omega\rangle   \big].
\eeqs
We refer to the three types according to their charge factors as uu, dd, and ud, respectively.
When fully contracted with the pion operator of $\left[\bar{u}\gamma_5 u -\bar{d} \gamma_5 d \right ]$ in Eq.\eqref{eq:op} at the source and sink, the decomposition can be written as,
\beqs
 \tilde{Q}_{\mu\mu}^{(PC)}(\bm q,t_3,t_2,t_1,t_0) &=\sum_{\bm x_2,\bm x_1} 
   e^{-i\bm q\cdot \bm x_2} e^{i\bm q\cdot \bm x_1} 
   \sum_{\bm x_3,\bm x_0} 
   \langle \Omega | \psi_{\pi^0}(\bm x_3,t_3)j^{(PC)}_{\mu}(\bm x_2,t_2)j^{(PC)}_{\mu}(\bm x_1,t_1)\psi_{\pi^0}^\dagger(\bm x_0,t_0)| \Omega\rangle \\
   &
=    \tilde{Q}_{\mu\mu}^{(uu)} +  \tilde{Q}_{\mu\mu}^{(dd)}  + \tilde{Q}_{\mu\mu}^{(ud)}.
   \label{eq:4pt}
\eeqs
The $uu$ type has 40 terms if u and d quarks are distinct. 
If the isospin limit is taken, we get 14 terms given by,
\beq
\tilde{Q}_{\mu\mu}^{(uu)} ={1\over 9}f^2 Z_V^2\kappa^2 \sum_{i=0}^{13} uu_i(\bm q,t_3,t_2,t_1,t_0),
\eeq
where
% \fontsize{9}{10}
\beqs
uu_{4}^{\text{A}}=4\,\text{tr}\left[S(t_1, t_3)\gamma_5S(t_3, t_2)\gamma_{\mu}e^{-i\mathbf{q}}S(t_2, t_0)\gamma_5S(t_0, t_1)\gamma_{\mu}e^{i\mathbf{q}}\right]\\
uu_{1}^{\text{A-bwd}}=4\,\text{tr}\left[S(t_2, t_3)\gamma_5S(t_3, t_1)\gamma_{\mu}e^{i\mathbf{q}}S(t_1, t_0)\gamma_5S(t_0, t_2)\gamma_{\mu}e^{-i\mathbf{q}}\right]\\
uu_{11}^{\text{B}}=4\,\text{tr}\left[S(t_2, t_3)\gamma_5S(t_3, t_0)\gamma_5S(t_0, t_1)\gamma_{\mu}e^{i\mathbf{q}}S(t_1, t_2)\gamma_{\mu}e^{-i\mathbf{q}}\right]\\
uu_{2}^{\text{B-bwd}}=4\,\text{tr}\left[S(t_0, t_3)\gamma_5S(t_3, t_2)\gamma_{\mu}e^{-i\mathbf{q}}S(t_2, t_1)\gamma_{\mu}e^{i\mathbf{q}}S(t_1, t_0)\gamma_5\right]\\
uu_{8}^{\text{C}}=4\,\text{tr}\left[S(t_1, t_3)\gamma_5S(t_3, t_0)\gamma_5S(t_0, t_2)\gamma_{\mu}e^{-i\mathbf{q}}S(t_2, t_1)\gamma_{\mu}e^{i\mathbf{q}}\right]\\
uu_{7}^{\text{C-bwd}}=4\,\text{tr}\left[S(t_0, t_3)\gamma_5S(t_3, t_1)\gamma_{\mu}e^{i\mathbf{q}}S(t_1, t_2)\gamma_{\mu}e^{-i\mathbf{q}}S(t_2, t_0)\gamma_5\right]\\
uu_{12}^{\text{D}}=-8\,\text{tr}\left[S(t_0, t_3)\gamma_5S(t_3, t_0)\gamma_5\right]\text{tr}\left[S(t_1, t_2)\gamma_{\mu}e^{-i\mathbf{q}}S(t_2, t_1)\gamma_{\mu}e^{i\mathbf{q}}\right]\\
uu_{10}^{\text{El}}=-4\,\text{tr}\left[S(t_1, t_3)\gamma_5S(t_3, t_0)\gamma_5S(t_0, t_1)\gamma_{\mu}e^{i\mathbf{q}}\right]\text{tr}\left[S(t_2, t_2)\gamma_{\mu}e^{-i\mathbf{q}}\right]\\
uu_{0}^{\text{El-bwd}}=-4\,\text{tr}\left[S(t_0, t_3)\gamma_5S(t_3, t_1)\gamma_{\mu}e^{i\mathbf{q}}S(t_1, t_0)\gamma_5\right]\text{tr}\left[S(t_2, t_2)\gamma_{\mu}e^{-i\mathbf{q}}\right]\\
uu_{9}^{\text{Er}}=-4\,\text{tr}\left[S(t_2, t_3)\gamma_5S(t_3, t_0)\gamma_5S(t_0, t_2)\gamma_{\mu}e^{-i\mathbf{q}}\right]\text{tr}\left[S(t_1, t_1)\gamma_{\mu}e^{i\mathbf{q}}\right]\\
uu_{5}^{\text{Er-bwd}}=-4\,\text{tr}\left[S(t_0, t_3)\gamma_5S(t_3, t_2)\gamma_{\mu}e^{-i\mathbf{q}}S(t_2, t_0)\gamma_5\right]\text{tr}\left[S(t_1, t_1)\gamma_{\mu}e^{i\mathbf{q}}\right]\\
uu_{13}^{\text{F}}=8\,\text{tr}\left[S(t_0, t_3)\gamma_5S(t_3, t_0)\gamma_5\right]\text{tr}\left[S(t_2, t_2)\gamma_{\mu}e^{-i\mathbf{q}}\right]\text{tr}\left[S(t_1, t_1)\gamma_{\mu}e^{i\mathbf{q}}\right]\\
uu_{3}^{\text{G}}=-4\,\text{tr}\left[S(t_2, t_3)\gamma_5S(t_3, t_2)\gamma_{\mu}e^{-i\mathbf{q}}\right]\text{tr}\left[S(t_0, t_1)\gamma_{\mu}e^{i\mathbf{q}}S(t_1, t_0)\gamma_5\right]\\
uu_{6}^{\text{H}}=-4\,\text{tr}\left[S(t_1, t_3)\gamma_5S(t_3, t_1)\gamma_{\mu}e^{i\mathbf{q}}\right]\text{tr}\left[S(t_0, t_2)\gamma_{\mu}e^{-i\mathbf{q}}S(t_2, t_0)\gamma_5\right].\\
\eeqs
\normalsize
The superscripts correspond to the topological diagrams depicted in Fig.~\ref{fig:4pt} and the subscripts are for book-keeping. The charge factors $q_u=2/3$ and $q_d=-1/3$ have been incorporated.
The first six terms in this equation are connected insertions (CI), the eight remaining ones are  disconnected insertions (DI).
The momentum factor is defined by a diagonal matrix,
\beq
[e^{\pm i{\bf q}}]_{s,c,{\bf x};s',c',{\bf x}'}\equiv 
\delta_{ss'}\delta_{cc'} \delta_{{\bf x},{\bf x'}}
e^{\pm i{\bf q}\cdot{\bf x}}.
\label{eq:mom}
\eeq
 The spatial sums over $(\bm x_2, \bm x_1, \bm x_3, \bm x_0)$ are implicit.

For $dd$ type we find a factor of 4 relative to $uu$ type for both CI and DI, term by term, so we write the relation as,
\beq
\tilde{Q}_{\mu\mu}^{(uu)} = 4 \, \tilde{Q}_{\mu\mu}^{(dd)}.
\eeq
This is a direct consequence of the charge factors $q_u^2=4/9$ vs. $q_d^2=1/9$. 

The crossed $ud$ type has 20 terms if u and d quarks are distinct, 7 in the isospin limit given by, 
\beq
\tilde{Q}_{\mu\mu}^{(ud)} ={1\over 9}f^2 Z_V^2\kappa^2 \sum_{i=0}^{6} ud_i(\bm q,t_3,t_2,t_1,t_0),
\eeq
where
% \fontsize{9}{10}
%
\beqs
ud_{6}^{\text{El}}=4\,\text{tr}\left[S(t_1, t_3)\gamma_5S(t_3, t_0)\gamma_5S(t_0, t_1)\gamma_{\mu}e^{i\mathbf{q}}\right]\text{tr}\left[S(t_2, t_2)\gamma_{\mu}e^{-i\mathbf{q}}\right] \\
ud_{4}^{\text{El-bwd}}=4\,\text{tr}\left[S(t_0, t_3)\gamma_5S(t_3, t_1)\gamma_{\mu}e^{i\mathbf{q}}S(t_1, t_0)\gamma_5\right]\text{tr}\left[S(t_2, t_2)\gamma_{\mu}e^{-i\mathbf{q}}\right]\\
ud_{1}^{\text{Er}}=4\,\text{tr}\left[S(t_2, t_3)\gamma_5S(t_3, t_0)\gamma_5S(t_0, t_2)\gamma_{\mu}e^{-i\mathbf{q}}\right]\text{tr}\left[S(t_1, t_1)\gamma_{\mu}e^{i\mathbf{q}}\right]\\
ud_{0}^{\text{Er-bwd}}=4\,\text{tr}\left[S(t_0, t_3)\gamma_5S(t_3, t_2)\gamma_{\mu}e^{-i\mathbf{q}}S(t_2, t_0)\gamma_5\right]\text{tr}\left[S(t_1, t_1)\gamma_{\mu}e^{i\mathbf{q}}\right]\\
ud_{5}^{\text{F}}=-8\,\text{tr}\left[S(t_0, t_3)\gamma_5S(t_3, t_0)\gamma_5\right]\text{tr}\left[S(t_2, t_2)\gamma_{\mu}e^{-i\mathbf{q}}\right]\text{tr}\left[S(t_1, t_1)\gamma_{\mu}e^{i\mathbf{q}}\right]\\
ud_{2}^{\text{G}}=-4\,\text{tr}\left[S(t_2, t_3)\gamma_5S(t_3, t_2)\gamma_{\mu}e^{-i\mathbf{q}}\right]\text{tr}\left[S(t_0, t_1)\gamma_{\mu}e^{i\mathbf{q}}S(t_1, t_0)\gamma_5\right]\\
ud_{3}^{\text{H}}=-4\,\text{tr}\left[S(t_1, t_3)\gamma_5S(t_3, t_1)\gamma_{\mu}e^{i\mathbf{q}}\right]\text{tr}\left[S(t_0, t_2)\gamma_{\mu}e^{-i\mathbf{q}}S(t_2, t_0)\gamma_5\right].\\
\eeqs
They are all disconnected insertions (digram D is absent). 

Taken together, the total for $\pi^0$ in the isospin limit is then given by 14 terms, 
\beq
\tilde{Q}_{\mu\mu}^{(PC)} ={1\over 9}f^2 Z_V^2\kappa^2 \sum_{i=0}^{13} g_i(\bm q,t_3,t_2,t_1,t_0),
\eeq
where
\beqs
g_{4}^{\text{A}}=5\,\text{tr}\left[S(t_1, t_3)\gamma_5S(t_3, t_2)\gamma_{\mu}e^{-i\mathbf{q}}S(t_2, t_0)\gamma_5S(t_0, t_1)\gamma_{\mu}e^{i\mathbf{q}}\right]\\
g_{1}^{\text{A-bwd}}=5\,\text{tr}\left[S(t_2, t_3)\gamma_5S(t_3, t_1)\gamma_{\mu}e^{i\mathbf{q}}S(t_1, t_0)\gamma_5S(t_0, t_2)\gamma_{\mu}e^{-i\mathbf{q}}\right]\\
g_{13}^{\text{B}}=5\,\text{tr}\left[S(t_2, t_3)\gamma_5S(t_3, t_0)\gamma_5S(t_0, t_1)\gamma_{\mu}e^{i\mathbf{q}}S(t_1, t_2)\gamma_{\mu}e^{-i\mathbf{q}}\right]\\
g_{2}^{\text{B-bwd}}=5\,\text{tr}\left[S(t_0, t_3)\gamma_5S(t_3, t_2)\gamma_{\mu}e^{-i\mathbf{q}}S(t_2, t_1)\gamma_{\mu}e^{i\mathbf{q}}S(t_1, t_0)\gamma_5\right]\\
g_{9}^{\text{C}}=5\,\text{tr}\left[S(t_1, t_3)\gamma_5S(t_3, t_0)\gamma_5S(t_0, t_2)\gamma_{\mu}e^{-i\mathbf{q}}S(t_2, t_1)\gamma_{\mu}e^{i\mathbf{q}}\right]\\
g_{7}^{\text{C-bwd}}=5\,\text{tr}\left[S(t_0, t_3)\gamma_5S(t_3, t_1)\gamma_{\mu}e^{i\mathbf{q}}S(t_1, t_2)\gamma_{\mu}e^{-i\mathbf{q}}S(t_2, t_0)\gamma_5\right]\\
g_{8}^{\text{D}}=-10\,\text{tr}\left[S(t_0, t_3)\gamma_5S(t_3, t_0)\gamma_5\right]\text{tr}\left[S(t_1, t_2)\gamma_{\mu}e^{-i\mathbf{q}}S(t_2, t_1)\gamma_{\mu}e^{i\mathbf{q}}\right]\\
g_{12}^{\text{El}}=-1\,\text{tr}\left[S(t_1, t_3)\gamma_5S(t_3, t_0)\gamma_5S(t_0, t_1)\gamma_{\mu}e^{i\mathbf{q}}\right]\text{tr}\left[S(t_2, t_2)\gamma_{\mu}e^{-i\mathbf{q}}\right]\\
g_{0}^{\text{El-bwd}}=-1\,\text{tr}\left[S(t_0, t_3)\gamma_5S(t_3, t_1)\gamma_{\mu}e^{i\mathbf{q}}S(t_1, t_0)\gamma_5\right]\text{tr}\left[S(t_2, t_2)\gamma_{\mu}e^{-i\mathbf{q}}\right]\\
g_{10}^{\text{Er}}=-1\,\text{tr}\left[S(t_2, t_3)\gamma_5S(t_3, t_0)\gamma_5S(t_0, t_2)\gamma_{\mu}e^{-i\mathbf{q}}\right]\text{tr}\left[S(t_1, t_1)\gamma_{\mu}e^{i\mathbf{q}}\right]\\
g_{5}^{\text{Er-bwd}}=-1\,\text{tr}\left[S(t_0, t_3)\gamma_5S(t_3, t_2)\gamma_{\mu}e^{-i\mathbf{q}}S(t_2, t_0)\gamma_5\right]\text{tr}\left[S(t_1, t_1)\gamma_{\mu}e^{i\mathbf{q}}\right]\\
g_{11}^{\text{F}}=2\,\text{tr}\left[S(t_0, t_3)\gamma_5S(t_3, t_0)\gamma_5\right]\text{tr}\left[S(t_2, t_2)\gamma_{\mu}e^{-i\mathbf{q}}\right]\text{tr}\left[S(t_1, t_1)\gamma_{\mu}e^{i\mathbf{q}}\right]\\
g_{3}^{\text{G}}=-9\,\text{tr}\left[S(t_2, t_3)\gamma_5S(t_3, t_2)\gamma_{\mu}e^{-i\mathbf{q}}\right]\text{tr}\left[S(t_0, t_1)\gamma_{\mu}e^{i\mathbf{q}}S(t_1, t_0)\gamma_5\right]\\
g_{6}^{\text{H}}=-9\,\text{tr}\left[S(t_1, t_3)\gamma_5S(t_3, t_1)\gamma_{\mu}e^{i\mathbf{q}}\right]\text{tr}\left[S(t_0, t_2)\gamma_{\mu}e^{-i\mathbf{q}}S(t_2, t_0)\gamma_5\right].\\
\label{eq:all}
\eeqs
We evaluate the first six terms (in their conserved current version) in this work; they are the connected insertions corresponding to diagrams A,B,C in Fig.~\ref{fig:4pt}. 
Details on how to enforce zero-momentum pions with wall-sources, utilization  of SST propagators, and implementation of conserved current can be found in Ref.~\cite{Lee:2023rmz} and  Ref.~\cite{Lee:2023lnx}.
%The remaining eight disconnected terms are computationally  challenging and are left for future work. 
%Their contributions to the total are expected to be sub-leading.

 %%%%%%%%%%%%%%
\end{widetext}
%%%%%%%%%%%%%%

\end{document}